\documentclass{article}

\usepackage{amsmath, amssymb, amsfonts, theorem}

\newcommand{\x}{\mathbf{x}}
\newcommand{\y}{\mathbf{y}}
\newcommand{\al}{\boldsymbol{\alpha}}
\newcommand{\bk}{\mathbf{k}}
\newcommand{\rxy}{|\x- \y|}
\newcommand{\D}{\displaystyle}
\newcommand{\rhs}{$\mathrm{r.\,h.\,s.}$ }

\newtheorem{lemma}{Lemma}
\newtheorem{theorem}{Theorem}
\newtheorem{remark}{Remark}
\newenvironment{proof}[1]{{\bf Proof{#1}.}}{$\bullet$}

\makeatletter
\@addtoreset{equation}{section}
\makeatother

\begin{document}

\title{Stability of Atoms in the Brown--Ravenhall Model}
\author{Sergey Morozov and Semjon Vugalter}

\maketitle

\begin{abstract}
We consider the Brown--Ravenhall model of a relativistic atom with $N$ electrons and a nucleus of charge $Z$ and prove the existence of an infinite number of discrete eigenvalues for $N\leqslant Z$. As an intermediate result we prove a HVZ--type theorem for these systems.
\end{abstract}

\section{Introduction}

A fundamental fact of quantum mechanics is that any stable state (a state which can exist infinitely long in time) of a quantum system corresponds to an eigenfunction of the Hamiltonian of this system. In particular, for an atom, which is known to be a stable system, this means that its lowest spectral point should be an eigenvalue. Hence the existence of an eigenvalue at the bottom of the spectrum of the Hamiltonian can be considered as one of the criteria of correctness of a mathematical model of an atom.

For multiparticle Schr\"odinger operators on the mathematically rigorous level the existence of discrete eigenvalues at the bottom of the spectrum was proved by T. Kato (for an atom with two electrons) \cite{Kato1951} and G. Zhislin (in the general case) \cite{Zhislin1960}. In the following 45 years these results were generalized in many different directions. The existence of ground states was proved for atoms in an external magnetic field (J. Avron, I. Herbst, B. Simon \cite{AvronHerbstSimon1978} and S. Vugalter, G. Zhislin \cite{VugalterZhislin1992}), for the Herbst operator (S. Vugalter, G. Zhislin \cite{VugalterZhislin2002}). The most recent development is the proof of the existence of a stable ground state in the Pauli--Fierz model, which describes an atom interacting with a quantized radiation field (V. Bach, J. Fr\"ohlich, I. Sigal \cite{BachFroehlichSigal1999}; J.--M. Barbaroux, T. Chen, S. Vugalter \cite{BarbarouxChenVugalter2003}; E. Lieb, M. Loss \cite{LiebLoss2003}).

In the work at hand we prove the existence of an infinite number of discrete eigenvalues, which accumulate at the bottom of the essential spectrum for the Brown--Ravenhall model of an atom or a positive atomic ion. The Brown--Ravenhall operator is one of the models used by physicists and quantum--chemists to describe relativistic effects in atoms (for a discussion of physical accuracy of this model see \cite{Sucher1980}).

The mathematically rigorous study of this operator started with the work by W. Evans, P. Perry and H. Siedentop \cite{EvansPerrySiedentop1996}, who proved that the Brown--Ravenhall operator of a one--particle Coulomb system with the nuclear charge $Z$ is semibounded from below for $\alpha Z\leqslant 2\big(\frac{\pi}{2}+ \frac{2}{\pi}\big)^{-1}$, where $\alpha$ is the fine structure constant. Further results on the lower bound of the spectrum of this operator with one electron and one or several nuclei were obtained by C. Tix \cite{Tix1997, Tix1998} and A. Balinsky, W. Evans \cite{BalinskyEvans1999}.

Although the questions of semiboundedness and self--adjointness of the\\ Brown--Ravenhall operator are extremely important, its successful application requires much more detailed knowledge of its spectral properties. The goal of this paper is to establish some of them.

A way to prove the existence of a stable ground state of a multiparticle system was developed by G. Zhislin in \cite{Zhislin1960} and consists of two steps. The first step is to prove a so--called HVZ--type theorem which establishes a criterion for the existence of a bound state. The second step is a construction of a trial function which satisfies this criterion. In the work at hand we follow the same strategy. First we prove a HVZ--type theorem (Theorem~\ref{HVZ theorem}), showing that the bottom of the essential spectrum of the Brown--Ravenhall operator of an atomic system with $N$ electrons is determined by the bottom of the spectrum of the operator with $N- 1$ electrons. Then we construct a trial function with an expectation value of the energy less than the bottom of the essential spectrum (Theorem~\ref{Z-theorem}). Although we use the same strategy as in the original paper by G. Zhislin, the work at hand is technically very different from the works on Schr\"odinger operators. The main differences are caused by the nonlocality of the Brown--Ravenhall operator. Notice that the HVZ theorem was proved earlier by R. Lewis, H. Siedentop and S. Vugalter \cite{LewisSiedentopVugalter1997} for the Herbst--type operator, which is also non--local. The Brown--Ravenhall operator in contrast to the Herbst operator has, however, not only a non--local kinetic energy, but also a non--local potential energy. This leads to a large number of additional complications.

In Section~\ref{facts} we prove several lemmata which allow us to estimate these non--local effects and to modify the method of \cite{Zhislin1960} in such a way that it works also for the Brown--Ravenhall operator. Here the most important role is played by the estimates on the commutator of the projector on the positive spectral subspace of the Dirac operator with a smoothed characteristic function of a region in the configuration space (Lemma~\ref{commutator lemma}) and the decay of the integral kernel of this projector (Lemma~\ref{decay lemma}).

When this work was in preparation, the authors received two preprints by D. Jakuba\ss a--Amundsen \cite{Doris1, Doris2}, where the HVZ theorem for the Brown--Ravenhall operator was proved in a different way and without taking the Pauli principle into account.

\section{Preliminaries}

The $N$--electron Brown--Ravenhall Hamiltonian is given by
\begin{equation}\label{H_N}
\mathcal{H}_N= \Lambda_+^N\bigg(\underset{n= 1}{\overset{N}{\sum}}\bigg(D_n-
\frac{\alpha Z}{|\x_n|}\bigg)+ \underset{n<
  j}{\overset{N}{\sum}}\frac{\alpha}{|\x_n- \x_j|}\bigg)\Lambda_+^N.
\end{equation}
Here $\x_n\in\mathbb{R}^3$ is the position vector of the $n^{th}$ electron. The Dirac operator in the standard representation is given by
\begin{equation}\label{Dirac}
D= -i\al\cdot\nabla+ \beta,
\end{equation}
where $\al:= (\alpha_1, \alpha_2, \alpha_3)$ and $\beta$ are the the four Dirac matrices, explicitly
\[
\alpha_k= \begin{pmatrix}0& \sigma_k\\ \sigma_k& 0\end{pmatrix}, \quad k= 1, 2, 3,
\]
$\sigma_k$ denoting the $k^{th}$ Pauli matrix
\[
\sigma_1=\begin{pmatrix}0& 1\\ 1& 0\end{pmatrix}, \quad \sigma_2=\begin{pmatrix}0& -i\\ i& 0\end{pmatrix}, \quad \sigma_3=\begin{pmatrix}1& 0\\ 0& -1\end{pmatrix},
\]
and
\[
\beta= \begin{pmatrix}1& 0& 0& 0\\ 0& 1& 0& 0\\ 0& 0& -1& 0\\ 0& 0& 0& -1\end{pmatrix}.
\]
$D_n$ is the Dirac operator $D$ acting on the coordinates of the $n^{th}$ electron. $Z$ is the nuclear charge, $\alpha\approx 1/137$ is the fine structure constant, and
\begin{equation}\label{Lambda^N}
\Lambda_+^N=\underset{n= 1}{\overset{N}{\otimes}}\Lambda_+^{(n)},
\end{equation}
where $\Lambda_+^{(n)}$ is the projector onto the positive spectral subspace of $D_n$. 

The underliying Hilbert space is $\mathfrak{H}^N:= \Lambda^N_+\underset{n= 1}{\overset{N}{\wedge}}L_2(\mathbb{R}^3, \mathbb{C}^4)$, where $\wedge$ stands for the antisymmetric tensor product of one--electron Hilbert spaces. The operator $\mathcal{H}_N$ is well defined as the operator corresponding to the semibounded closed form on $\Lambda_+^N\underset{n= 1}{\overset{N}{\wedge}}H^{1/2}(\mathbb{R}^3, \mathbb{C}^4)$ for
\begin{equation}\label{Z_c}
\alpha Z< \alpha Z_c= \frac{2}{\big(\frac{\pi}{2}+ \frac{2}{\pi}\big)}.
\end{equation}
We always assume in the following that condition \eqref{Z_c} is fulfilled.

The semiboundedness of $\mathcal{H}_N$ follows from the semiboundedness of the one--particle operators $\Lambda_+\Big(D- \frac{\D\alpha Z}{\D|\x|}\Big)\Lambda_+$ (see \cite{EvansPerrySiedentop1996}) and the positivity of the two--particle interaction terms.

We denote the spectrum of an arbitrary selfadjoint operator $A$ by $\sigma(A)$. $[A, B]= AB- BA$ is the commutator of two operators. $B(R, \x)$ is the open ball in $\mathbb{R}^d$ of radius $R> 0$ centered at $\x$. $B(R):= B(R, \mathbf{0})$. $\langle\cdot, \cdot\rangle$ and $\|\cdot\|$ stand for the inner product and the norm in $L_2(\mathbb{R}^{3d}, \mathbb{C}^{4^d})$. $d$ is usually clear by context. Let $E_{N- 1}:= \inf \sigma(\mathcal{H}_{N- 1})$. Irrelevant constants are denoted by $C$. $I_\Omega$ is the indicator function of the set $\Omega$. The Fourier transform of $f$ is denoted by $\hat f$.

In auxiliary calculations it is sometimes convenient to consider the operator \eqref{H_N} in the space $\Lambda^N_+\underset{n= 1}{\overset{N}{\otimes}}L_2(\mathbb{R}^3, \mathbb{C}^4)$, i. e. without antisymmetrization. We use this extension without changing the notation.

The main result of this article are the following two theorems.

\begin{theorem}\label{HVZ theorem}
For any $N> 1$, we have
\begin{equation}\label{HVZ}
\sigma_{\mathrm{ess}}(\mathcal{H}_N)= [E_{N- 1}+ 1, \infty).
\end{equation}
\end{theorem}

\begin{theorem}\label{Z-theorem}
Let $N\leqslant Z$. Then the operator $\mathcal{H}_N$ has infinitely many eigenvalues below the essential spectrum.
\end{theorem}

\begin{remark}
Theorem \ref{HVZ theorem} is an analogue of the HVZ theorem for multiparticle Schr\"o\-din\-ger operators (see \cite{CyconFroeseKirschSimon1987} and the original papers \cite{Zhislin1960, vanWinter1964, Hunziker1966}). Analogous theorems were proved for the magnetic Schr\"odinger operator \cite{VugalterZhislin1991} and the Herbst operator \cite{LewisSiedentopVugalter1997}.
\end{remark}

\begin{remark}
In contrast to the Schr\"odinger case the bottom of the essential spectrum of $\mathcal{H}_N$ is $E_{N- 1}+ 1$ and not $E_N$. This is related to the fact that in the Brown--Ravenhall model the spectrum of the free electron is $[1, \infty)$ instead of $[0, \infty)$.
\end{remark}

\begin{remark}
In the multiparticle Schr\"odinger case the existence of discrete eigenvalues was proved in \cite{Kato1951} for $N= 2$, and in \cite{Zhislin1960} for arbitrary $N$.
\end{remark}

The proof of Theorem \ref{HVZ theorem} is given in Sections \ref{easy part} and \ref{hard part}. Theorem \ref{Z-theorem} is proved in Section \ref{existence of eigenvalues}. Section \ref{facts} contains some lemmata used in the subsequent sections.

\section{Technical Lemmata}\label{facts}

\subsection{Commutator Estimate}

The projector $\Lambda_+$ for the free one--particle Dirac operator is given by (see \cite{Thaller1992}, formula 1.1.54)
\begin{equation}\label{Lambda_+}
\Lambda_+=  \frac{1}{2}+ \frac{D}{2|D|}= \frac{1}{2}+
\mathcal{F}^*\frac{\al\cdot\mathbf{p}+ \beta}{2\sqrt{|\mathbf{p}|^2+
    1}}\mathcal{F},
\end{equation}
where $\mathcal{F}$ is the Fourier transform. In the coordinate representation for\\ $f\in C_0^1(\mathbb{R}^3, \mathbb{C}^4)$ the operator $\Lambda_+$ acts as
\begin{equation}\begin{split}\label{Lambda_+ in configuration space}
(\Lambda_+f)(\x)= \frac{f(\x)}{2}+
\frac{1}{4\pi^2}\int\limits_{\mathbb{R}^3}\bigg(\beta\frac{K_1\big(|\x- \y|\big)}{|\x- \y|}+
\frac{i\al\cdot(\mathbf{x}- \mathbf{y})}{|\x-
  \y|^2}K_0\big(\rxy\big)\bigg)f(\y)d\y\\+ \frac{i}{2\pi^2}\underset{\varepsilon\rightarrow +0}{\lim}\int\limits_{\mathbb{R}^3\setminus B(\varepsilon, \x)}\frac{\al\cdot(\x- \y)}{\rxy^3}K_1\big(\rxy\big)f(\y)d\y,
\end{split}\end{equation}
where the limit on the \rhs is the limit in $L_2(\mathbb{R}^3, \mathbb{C}^4)$ (see Appendix~\ref{integral formula}).
For convenience of the reader we state properties of the functions $K_\nu,\: \nu= 1, 2$, in Appendix~\ref{K-functions}.

\begin{lemma}\label{commutator lemma} 
Let $\chi\in C^2(\mathbb{R}^3)$. Then the norm of the operator 
\[
[\chi, \Lambda_+]:L_2(\mathbb{R}^3, \mathbb{C}^4)\rightarrow H^1(\mathbb{R}^3, \mathbb{C}^4)
\]
satisfies
\begin{equation}\label{commutator}
\big\|[\chi, \Lambda_+]\big\|_{L_2(\mathbb{R}^3, \mathbb{C}^4)\rightarrow
H^1(\mathbb{R}^3, \mathbb{C}^4)}\leqslant C\big(\|\nabla\chi\|_\infty+ \|\partial^2\chi\|_\infty\big). 
\end{equation}
Here $\|\partial^2\chi\|_\infty= \underset{\substack{\mathbf{z}\in\mathbb{R}^3\\ k,l\in\{1, 2,
    3\}}}{\max}\big|\partial_{kl}^2\chi(\mathbf{z})\big|$.
\end{lemma}

In the proof of Lemma~\ref{commutator lemma} we shall apply the following theorem, which we formulate here for convenience of the reader.

\begin{theorem}\label{Stein theorem}
{\bfseries\em (Stein \cite{Stein1970}, Chapter~2, sec.~3.2)} Let $K:\mathbb{R}^n\rightarrow\mathbb{C}$ be measurable such that
\begin{equation}\label{condition 1}
\big|K(\x)\big|\leqslant B|\x|^{-n},\quad \x\neq \mathbf{0},
\end{equation}
\begin{equation}\label{condition 2}
\int\limits_{|\x|\geqslant 2|\y|}\big|K(\x- \y)- K(\x)\big|d^n\x\leqslant B,\quad 0< |\y|,
\end{equation}
and
\begin{equation}\label{condition 3}
\int\limits_{R_1< |\x|< R_2}K(\x)d^n\x= 0, \quad\text{for all}\quad 0< R_1< R_2< \infty.
\end{equation}
For an arbitrary $f\in L_p(\mathbb{R}^n),\: 1< p< \infty$, let
\begin{equation}\label{convolution}
T_\varepsilon(f)(\x)= \int\limits_{|\y|\geqslant \varepsilon}f(\x- \y)K(\y)d^n\y,\quad \varepsilon> 0.
\end{equation}
Then
\begin{equation}\label{boundedness}
\big\|T_\varepsilon(f)\big\|_p\leqslant A_p\|f\|_p
\end{equation}
with $A_p$ independent of $f$ and $\varepsilon$.
\end{theorem}
\begin{remark}
Inequality \eqref{boundedness} shows that the operator $T= \underset{\varepsilon\rightarrow +0}{\lim}T_\varepsilon$ exists as a bounded operator in $L_p(\mathbb{R}^n)$ and its norm satisfies $\|T\|_p\leqslant A_p$.
\end{remark}
\begin{proof}{ of Lemma~\ref{commutator lemma}}
Let us first prove that $[\chi, \Lambda_+]$ is a bounded operator in $L_2(\mathbb{R}^3, \mathbb{C}^4)$.
For $f\in C_0^1(\mathbb{R}^3, \mathbb{C}^4)$ formula \eqref{Lambda_+ in configuration space} implies
\begin{equation}\begin{split}\label{1}
\big([\chi, \Lambda_+]f\big)(\x)&\\= \frac{1}{4\pi^2}\int\bigg(&\beta\frac{K_1\big(|\x- \y|\big)}{|\x- \y|}+
\frac{i\al\cdot(\mathbf{x}- \mathbf{y})}{|\x-
  \y|^2}K_0\big(\rxy\big)\bigg)\big(\chi(\x)- \chi(\y)\big)f(\y)d\y\\&+
\frac{i}{2\pi^2}\int\frac{\al\cdot(\x- \y)}{\rxy^3}K_1\big(\rxy\big)\big(\chi(\x)- \chi(\y)\big)f(\y)d\y.
\end{split}\end{equation}

Estimating $\big|\chi(\x)- \chi(\y)\big|$ by $\rxy\|\nabla\chi\|_\infty$, using the density of $C_0^1(\mathbb{R}^3, \mathbb{C}^4)$ in $L_2(\mathbb{R}^3, \mathbb{C}^4)$ and applying the Young inequality for the convolution with a kernel from $L_1(\mathbb{R}^3)$, we arrive at
\begin{equation}\begin{split}\label{2}
\big\|[\chi&, \Lambda_+]\big\|_{L_2(\mathbb{R}^3, \mathbb{C}^4)\rightarrow L_2(\mathbb{R}^3, \mathbb{C}^4)}\\ &\leqslant \|\nabla\chi\|_\infty
\frac{1}{4\pi^2}\int\bigg(K_1\big(|\x|\big)+
3K_0\big(|\x|\big)+ 6\frac{K_1\big(|\x|\big)}{|\x|}\bigg)d\x\leqslant C\|\nabla\chi\|_\infty.
\end{split}\end{equation}

To complete the proof of Lemma \ref{commutator lemma} one has to show that
\[
\big\|\nabla[\chi, \Lambda_+]f\big\|_{L_2(\mathbb{R}^3, \mathbb{C}^4)}\leqslant C\big(\|\nabla\chi\|_\infty+ \|\partial^2\chi\|_\infty\big)\|f\|_{L_2(\mathbb{R}^3, \mathbb{C}^4)}.
\]
The differentiation of the first summand on the \rhs of \eqref{1} gives absolutely convergent integrals whose $L_2$--norms are bounded by $C\|\nabla\chi\|_\infty\|f\|$. The differentiation of the second integral on the \rhs of \eqref{1} in the $j^{th}$ component of $\x$ gives for $f\in C_0^1(\mathbb{R}^3, \mathbb{C}^4)$ (cf. Appendix~\ref{integral formula})
\begin{equation}\begin{split}\label{3}
\frac{i}{2\pi^2}\underset{\varepsilon\rightarrow +0}{\lim}\int\limits_{\mathbb{R}^3\setminus B(\varepsilon)}\bigg(\alpha_j\frac{K_1\big(\rxy\big)}{\rxy^3}- \frac{\al\cdot(\x-
  \y)(x_j- y_j)K_0\big(\rxy\big)}{\rxy^4}\\- \frac{4\al\cdot(\x- \y)(x_j-
  y_j)K_1\big(\rxy\big)}{\rxy^5}\bigg)\big(\chi(\x)- \chi(\y)\big)f(\y)d\y\\+
  \frac{i}{2\pi^2}\underset{\varepsilon\rightarrow +0}{\lim}\int\limits_{\mathbb{R}^3\setminus B(\varepsilon)}\frac{\al\cdot(\x-
  \y)K_1\big(\rxy\big)}{\rxy^3}\partial_j\chi(\x)f(\y)d\y.
\end{split}\end{equation}
The term 
\[
-\frac{i}{2\pi^2}\int\frac{\al\cdot(\x- \y)(x_j- y_j)K_0\big(\rxy\big)}{\rxy^4}\big(\chi(\x)- \chi(\y)\big)f(\y)d\y
\]
is bounded by $C\|\nabla\chi\|_\infty\|f\|$.

The Taylor expansion of $\chi$ gives
\begin{equation}\label{Taylor}
\chi(\x)- \chi(\y)= (x_k- y_k)\partial_k\chi(\x)- \frac{1}{2}(x_k-
y_k)(x_l- y_l)\partial_{kl}^2\chi(\mathbf{z}_{xy}), \quad
\mathbf{z}_{xy}\in [\x, \y].
\end{equation}
Here $[\x, \y]$ is the line segment connecting $\x$ and $\y$ and
summations in $k$ and $l$ from $1$ to $3$ are assumed. Substituting \eqref{Taylor} into \eqref{3} we arrive at
\begin{equation}\begin{split}\label{4}
\frac{i\partial_k\chi(\x)}{2\pi^2}\underset{\varepsilon\rightarrow +0}{\lim}\int\limits_{\mathbb{R}^3\setminus B(\varepsilon)}\bigg(\alpha_j- \frac{4\al\cdot(\x- \y)(x_j- y_j)}{\rxy^2}&\bigg)\frac{K_1\big(\rxy\big)(x_k- y_k)}{\rxy^3}f(\y)d\y\\+ \frac{i\partial_j\chi(\x)}{2\pi^2}\underset{\varepsilon\rightarrow +0}{\lim}\int\limits_{\mathbb{R}^3\setminus B(\varepsilon)}\frac{\al\cdot(\x- \y)K_1\big(\rxy\big)}{\rxy^3}&f(\y)d\y\\- \frac{i}{4\pi^2}\int\bigg(\alpha_j- \frac{4\al\cdot(\x- \y)(x_j- y_j)}{\rxy^2}&\bigg)\frac{K_1\big(\rxy\big)}{\rxy^3}\\ \times(x_k&- y_k)(x_l- y_l)\partial_{kl}^2\chi(\mathbf{z}_{xy})f(\y)d\y.
\end{split}\end{equation}

To the first two integrals in \eqref{4} we shall apply Theorem~\ref{Stein theorem} with $n= 3$.
Hypotheses \eqref{condition 1} and \eqref{condition 3} are obviously fulfilled. To prove that condition~\eqref{condition 2} is also fulfilled we note that \eqref{condition 2} follows from the estimate
\begin{equation}\label{condition 2'}
\big|\nabla K(\x)\big|\leqslant B|\x|^{-4}
\end{equation}
(see \cite{Stein1970}, page 34). Inequality \eqref{condition 2'} follows from \eqref{derivatives}. Due to Theorem~\ref{Stein theorem} the $L_2$--norms of the first two integrals in \eqref{4} are bounded by $C\|f\|_{L_2(\mathbb{R}^3, \mathbb{C}^4)}$. The last term in \eqref{4} is a convolution with a function from $L_1(\mathbb{R}^3)$, whose $L_2$--norm due to the Young inequality can be estimated with
\[
C\|\partial^2\chi\|_\infty\|f\|_{L_2(\mathbb{R}^3, \mathbb{C}^4)}.
\]
This completes the proof of Lemma~\ref{commutator lemma}.
\end{proof}

\subsection{Non--local Properties of the Operator $\Lambda_+$}

\begin{lemma}\label{decay lemma}
Let $\mathrm{supp} f\subset\Omega\subset\mathbb{R}^3, \: |\Omega|<\infty, \: \x\in\mathbb{R}^3, \: d:= \mathrm{dist}(\x, \Omega)> 0$. Then
\begin{equation}\label{decay}
\big|(\Lambda_+ f)(\x)\big|\leqslant G(d)|\Omega|^{1/2}\|f\|_{L_2(\mathbb{R}^3, \mathbb{C}^4)},
\end{equation}
where
\begin{equation}\label{G(d)}
G(d)= \frac{1}{4\pi^2}\Big(\frac{K_1(d)}{d}+ 3\frac{K_0(d)}{d}+ 6\frac{K_1(d)}{d^2}\Big).
\end{equation}
\end{lemma}
\begin{proof}{}
The statement of Lemma~\ref{decay lemma} follows immediately from the Schwarz inequality and formula \eqref{Lambda_+ in configuration space}, if we note that for $\x\notin\mathrm{supp}\,f$ all integrals in \eqref{Lambda_+ in configuration space} converge absolutely.
\end{proof}

\begin{remark}
Notice that the functions $K_\nu(d), \: \nu= 0,\,1$, and consequently $G(d)$, decay exponentially according to \eqref{asymptotics} as $d\rightarrow \infty$. Our proof of the HVZ theorem and of the existence of the discrete spectrum follows the same lines as in the original paper by G. Zhislin~\cite{Zhislin1960} for the Schr\"odinger operator. The new obstacle which was overcome in the present work is the non--locality of the Brown--Ravenhall operator. Lemma~\ref{decay lemma} tells us that, although the operator $\Lambda_+$ is non--local, for a compactly supported function $f$ the function $\Lambda_+f$ decays exponentially with the distance to the support of $f$.
\end{remark}

\subsection{Localization Error Estimate}

\begin{lemma}\label{multiplicator lemma}
Any bounded function $\chi\in C^1(\mathbb{R}^d)$ with bounded derivatives is a multiplicator in $H^{1/2}(\mathbb{R}^d, \mathbb{C}^k)$ for any $d, k\in\mathbb{N}$:
\begin{equation}\label{multiplier}
\|\chi u\|_{H^{1/2}(\mathbb{R}^d, \mathbb{C}^k)}\leqslant C_d\cdot\big(\|\chi\|_{L_\infty(\mathbb{R}^d)}+ \|\nabla\chi\|_{L_\infty(\mathbb{R}^d)}\big)\|u\|_{H^{1/2}(\mathbb{R}^d, \mathbb{C}^k)},
\end{equation}
for all $u\in H^{1/2}(\mathbb{R}^d, \mathbb{C}^k)$.
\end{lemma}
\begin{proof}{ of Lemma~\ref{multiplicator lemma}}
We choose the norm in $H^{1/2}(\mathbb{R}^d, \mathbb{C}^k)$ as (see \cite{Adams1975}, Theorem 7.48).
\begin{equation}\label{the norm}
\|u\|_{H^{1/2}(\mathbb{R}^d, \mathbb{C}^k)}^2:= \|u\|_{L_2(\mathbb{R}^d, \mathbb{C}^k)}^2+ \iint\frac{\big|u(\x)- u(\y)\big|^2}{|\x- \y|^{d+ 1}}d\x d\y.
\end{equation}
Then
\begin{equation}\label{chain}\begin{split}
\|\chi u\|_{H^{1/2}(\mathbb{R}^d, \mathbb{C}^k)}^2&= \|\chi u\|_{L_2(\mathbb{R}^d, \mathbb{C}^k)}^2+ \iint\frac{\big|\chi(\x)u(\x)- \chi(\y)u(\y)\big|^2}{|\x- \y|^{d+ 1}}d\x d\y\\ \leqslant \|\chi\|_{L_\infty}^2\|u\|_{L_2}^2&+ \iint\bigg(\frac{\big|\chi(\x)\big|^2\big|u(\x)- u(\y)\big|^2}{|\x- \y|^{d+ 1}}+ \frac{\big|\chi(\x)- \chi(\y)\big|^2\big|u(\y)\big|^2}{|\x- \y|^{d+ 1}}\bigg)d\x d\y\\ &\leqslant \|\chi\|_{L_\infty}^2\|u\|_{H^{1/2}}^2+ \underset{\y\in\mathbb{R}^d}{\sup}\int\frac{\big|\chi(\x)- \chi(\y)\big|^2}{|\x- \y|^{d+ 1}}d\x\|u\|_{L_2}^2.
\end{split}\end{equation}
The supremum on the \rhs of \eqref{chain} can be estimated as
\begin{equation}\label{estimate for integral}\begin{split}
&\underset{\y\in\mathbb{R}^d}{\sup}\int\frac{\big|\chi(\x)- \chi(\y)\big|^2}{|\x- \y|^{d+ 1}}d\x\leqslant \underset{\y\in\mathbb{R}^d}{\sup}\int\limits_{B(1, \y)}\frac{\big|\chi(\x)- \chi(\y)\big|^2}{|\x- \y|^{d+ 1}}d\x\\ &+ \underset{\y\in\mathbb{R}^d}{\sup}\int\limits_{\mathbb{R}^d\setminus B(1, \y)}\frac{\big|\chi(\x)- \chi(\y)\big|^2}{|\x- \y|^{d+ 1}}d\x\leqslant |S_{d- 1}|\big(\|\nabla\chi\|_{L_\infty}^2+ 4\|\chi\|_{L_\infty}^2\big),
\end{split}\end{equation}
where $|S_{d- 1}|$ is the area of $(d- 1)$--dimensional unit sphere. Substituting \eqref{estimate for integral} into \eqref{chain} and using the inequality $a^2+ b^2\leqslant (a+ b)^2$ for $a, b\geqslant 0$ we obtain \eqref{multiplier}.
\end{proof}

\begin{lemma}\label{boundedness of many-particle commutator in H^1/2}
For any bounded function $\chi\in C^1(\mathbb{R}^{3N})$ with bounded derivatives the operator $[\chi, \Lambda_+^N]$ is bounded in $H^{1/2}(\mathbb{R}^{3N}, \mathbb{C}^{4^N})$ and for any $\psi\in H^{1/2}(\mathbb{R}^{3N}, \mathbb{C}^{4^N})$ we have
\begin{equation}\label{jjj}\begin{split}
&\big\|[\chi, \Lambda_+^N]\psi\big\|_{H^{1/2}(\mathbb{R}^{3N}, \mathbb{C}^{4^N})}\\ &\leqslant C_N\big(\|\nabla\chi\|_{L_\infty(\mathbb{R}^{3N})}+ \|\partial^2\chi\|_{L_\infty(\mathbb{R}^{3N})}\big)\big(1+ \|\nabla\chi\|_{L_\infty(\mathbb{R}^{3N})}\big)\|\psi\|_{H^{1/2}(\mathbb{R}^{3N}, \mathbb{C}^{4^N})}.
\end{split}\end{equation}
\end{lemma}
\begin{proof}{}
Successively commuting $\chi$ with one--particle projections $\Lambda_+^{(n)}, \, n= 1, \dots, N$ (see \eqref{Lambda^N}) we obtain
\begin{equation}\label{reduction to one particle}
[\chi, \Lambda_+^N]= \sum_{n= 1}^N\prod_{k= 1}^{n- 1}\Lambda_+^{(k)}[\chi, \Lambda_+^{(n)}]\prod_{l= n+ 1}^N\Lambda_+^{(l)},
\end{equation}
where the empty products should be replaced by identity operators.
According to \eqref{Lambda_+}, the operator $\Lambda_+$ is bounded in $H^s(\mathbb{R}^3, \mathbb{C}^4)$ for any $s\in\mathbb{R}$. This, together with \eqref{reduction to one particle}, and Lemmata~\ref{commutator lemma} and \ref{multiplicator lemma}, implies \eqref{jjj}. 
\end{proof}

\begin{lemma}\label{control lemma}
There exists a constant $C_{N, Z}$ depending on $N$ and $Z$ such that for any $\psi\in\Lambda^N_+\underset{n= 1}{\overset{N}{\wedge}}H^{1/2}(\mathbb{R}^3, \mathbb{C}^4)$
\begin{equation}\label{control}
\langle\mathcal{H}_N\psi, \psi\rangle\geqslant C_{N, Z}\|\psi\|_{H^{1/2}(\mathbb{R}^{3N}, \mathbb{C}^{4^N})}^2.
\end{equation}
\end{lemma}
\begin{proof}{}
For the one particle Brown--Ravenhall operator one has (see \cite{Tix1998})
\begin{equation}\label{semiboundedness}
\Lambda_+\Big(D- \frac{\alpha Z}{|\x|}\Big)\Lambda_+\geqslant (1- \alpha Z)\Lambda_+.
\end{equation}
Inequality \eqref{semiboundedness} holds true for any $Z\leqslant Z_c$ (cf. \eqref{Z_c}).
Using \eqref{semiboundedness} with $Z= Z_c$ we get
\begin{equation}\label{without Coulomb}
\Lambda_+\Big(D- \frac{\alpha Z}{|\x|}\Big)\Lambda_+\geqslant \frac{Z_c- Z}{Z_c}D\Lambda_++ \frac{Z}{Z_c}(1- \alpha Z_c)\Lambda_+\geqslant \frac{Z_c- Z}{Z_c}D\Lambda_+.
\end{equation}
Now, since $\sum_{n< j}^N\dfrac{\alpha}{|\x_n- \x_j|}>0$, $Z< Z_c$, and $\psi= \Lambda_+^N\psi$, using \eqref{without Coulomb} we obtain
\begin{equation}\label{H^1/2 norm via H_N}\begin{split}
\langle\mathcal{H}_N\psi, \psi\rangle\geqslant \langle\underset{n= 1}{\overset{N}{\sum}}\Big(D_n- \frac{\alpha Z}{|\x_n|}\Big)\psi, \psi\rangle\geqslant C_Z\langle\underset{n= 1}{\overset{N}{\sum}}D_n\psi, \psi\rangle\geqslant C_{N, Z}\|\psi\|_{H^{1/2}(\mathbb{R}^{3N}, \mathbb{C}^{4^N})}^2,
\end{split}\end{equation}
which is \eqref{control}
\end{proof}

\begin{lemma}\label{localization lemma}
Let $\{\chi_a\}_{a\in\mathcal{A}}$ be a partition of unity with the properties 
\begin{equation}\label{partition requirements}
\chi_a\in C^2(\mathbb{R}^{3N}), \quad
\chi_a\geqslant 0, \quad \underset{a\in\mathcal{A}}{\sum}\chi_a^2= 1. 
\end{equation}
Then for any $\psi\in\Lambda_+^N\underset{n= 1}{\overset{N}{\wedge}}H^{1/2}(\mathbb{R}^3, \mathbb{C}^4)$ we have
\begin{equation}\label{localization formula}\begin{split}
&\bigg|\langle \mathcal{H}_N\psi, \psi\rangle-
\underset{a\in\mathcal{A}}{\sum}\langle \mathcal{H}_N\Lambda_+^N\chi_a\psi, \Lambda_+^N\chi_a\psi\rangle\bigg|\\ &\leqslant
\widetilde C_{N, Z}\sum_{a\in\mathcal{A}}\big(\|\nabla\chi_a\|_{L_\infty(\mathbb{R}^{3N})}+ \|\partial^2\chi_a\|_{L_\infty(\mathbb{R}^{3N})}\big)\big(1+ \|\nabla\chi_a\|_{L_\infty(\mathbb{R}^{3N})}\big)^2\langle\mathcal{H}_N\psi, \psi\rangle.
\end{split}\end{equation}
\end{lemma}
\begin{proof}{}
We write
\begin{equation}\begin{split}\label{localization calculations}
\langle \mathcal{H}_N\psi&, \psi\rangle= \langle\bigg(\underset{n= 1}{\overset{N}{\sum}}\Big(D_n-
\frac{\alpha Z}{|\x_n|}\Big)+ \underset{n<
  j}{\overset{N}{\sum}}\frac{\alpha}{|\x_n-
  \x_j|}\bigg)\underset{a\in\mathcal{A}}{\sum}\chi_a^2\Lambda_+^N\psi,
\Lambda_+^N\psi\rangle\\&= \underset{a\in\mathcal{A}}{\sum}\langle\bigg(\underset{n= 1}{\overset{N}{\sum}}\Big(D_n-
\frac{\alpha Z}{|\x_n|}\Big)+ \underset{n<
  j}{\overset{N}{\sum}}\frac{\alpha}{|\x_n-
  \x_j|}\bigg)\chi_a\Lambda_+^N\psi,
\chi_a\Lambda_+^N\psi\rangle\\&= \underset{a\in\mathcal{A}}{\sum}\langle\mathcal{H}_N\Lambda_+^N\chi_a\psi,
\Lambda_+^N\chi_a\psi\rangle\\&+ \underset{a\in\mathcal{A}}{\sum}\langle\bigg(\underset{n= 1}{\overset{N}{\sum}}\Big(D_n-
\frac{\alpha Z}{|\x_n|}\Big)+ \underset{n<
  j}{\overset{N}{\sum}}\frac{\alpha}{|\x_n-
  \x_j|}\bigg)[\chi_a, \Lambda_+^N]\psi,
\chi_a\Lambda_+^N\psi\rangle\\&+
\underset{a\in\mathcal{A}}{\sum}\langle\Lambda_+^N\chi_a\psi, \bigg(\underset{n= 1}{\overset{N}{\sum}}\Big(D_n-
\frac{\alpha Z}{|\x_n|}\Big)+ \underset{n<
  j}{\overset{N}{\sum}}\frac{\alpha}{|\x_n-
  \x_j|}\bigg)[\chi_a, \Lambda_+^N]\psi\rangle.
\end{split}\end{equation}
In the second line of \eqref{localization calculations} we used the relation
\begin{equation}\label{first order commutation}
\underset{a\in\mathcal{A}}{\sum}\langle g, \nabla(\chi_a^2 g)\rangle= \underset{a\in\mathcal{A}}{\sum}\langle\chi_a g, \nabla(\chi_a g)\rangle+ \underset{a\in\mathcal{A}}{\sum}\langle g, \nabla\Big(\frac{\chi_a^2}{2}\Big)g\rangle,
\end{equation}
which holds for any $g\in H^{1/2}(\mathbb{R}^3, \mathbb{C}^4)$. The last term in \eqref{first order commutation} is zero because of \eqref{partition requirements}.

It remains to estimate the last two terms on the \rhs of \eqref{localization calculations}.
Since the sesquilinear form of $\underset{n= 1}{\overset{N}{\sum}}\Big(D_n- \dfrac{\alpha Z}{|\x_n|}\Big)+ \underset{n< j}{\overset{N}{\sum}}\dfrac{\alpha}{|\x_n- \x_j|}$ is bounded on $H^{1/2}(\mathbb{R}^{3N}, \mathbb{C}^{4^N})$ (for the potential energy terms this follows from the Kato inequality), Lemmata~\ref{multiplicator lemma} and \ref{boundedness of many-particle commutator in H^1/2} imply that 
\begin{equation}\label{last terms}\begin{split}
\bigg|\langle\bigg(\underset{n= 1}{\overset{N}{\sum}}\Big(D_n&- \frac{\alpha Z}{|\x_n|}\Big)+ \underset{n< j}{\overset{N}{\sum}}\frac{\alpha}{|\x_n- \x_j|}\bigg)[\chi_a, \Lambda_+^N]\psi, \chi_a\Lambda_+^N\psi\rangle\\&+ \langle\Lambda_+^N\chi_a\psi, \bigg(\underset{n= 1}{\overset{N}{\sum}}\Big(D_n- \frac{\alpha Z}{|\x_n|}\Big)+ \underset{n< j}{\overset{N}{\sum}}\frac{\alpha}{|\x_n- \x_j|}\bigg)[\chi_a, \Lambda_+^N]\psi\rangle\bigg|\\ \leqslant \widetilde C_{N, Z}\big(\|\nabla&\chi_a\|_{L_\infty(\mathbb{R}^{3N})}+ \|\partial^2\chi_a\|_{L_\infty(\mathbb{R}^{3N})}\big)\\&\times\big(1+ \|\nabla\chi_a\|_{L_\infty(\mathbb{R}^{3N})}\big)^2\|\psi\|_{H^{1/2}(\mathbb{R}^{3N}, \mathbb{C}^{4^N})}^2,\quad\textrm{for all}\quad a\in\mathcal{A}.
\end{split}\end{equation}
The relation \eqref{localization formula} follows from \eqref{localization calculations}, \eqref{last terms} and Lemma~\ref{control lemma}.
\end{proof}

\section{Proof of Theorem~\ref{HVZ theorem}: ``Easy Part''}\label{easy part}

We shall prove that
\begin{equation}\label{easy claim}
\sigma_{\mathrm{ess}}(\mathcal{H}_N)\supseteq [E_{N- 1}+ 1, \infty),
\end{equation}
by construction of an appropriate Weyl sequence for $\mathcal{H}_N$ at the point $E_{N- 1}+ \lambda$ for any $\lambda\geqslant 1$.

Since $\mathcal{H}_N$ commutes with the projector $P_A$ onto the antisymmetric subspace, it suffices to find a Weyl sequence $\{\Psi_l\}_{l= 1}^\infty\in\underset{n= 1}{\overset{N}{\otimes}}L_2(\mathbb{R}^3, \mathbb{C}^4)$ such that for $l$ big enough
\begin{equation}\label{lower norm bound}
\|P_A\Psi_l\|> \delta_0, \quad \delta_0> 0,
\end{equation}
where $\delta_0$ is independent of $l$.

For $j\in\mathbb{N}$ let
\begin{equation}\label{varphi_j}
\varphi_j\in P_{[E_{N- 1}, E_{N- 1}+ j^{-1})}(\mathcal{H}_{N- 1})\mathfrak{H}^{N- 1}, \quad \|\varphi_j\|_{L_2(\mathbb{R}^{3(N- 1)}, \mathbb{C}^{4^{(N- 1)}})}= 1.
\end{equation}
Here $P_J(\mathcal{H}_{N- 1})$ is the spectral projector of $\mathcal{H}_{N- 1}$ corresponding to the interval $J$.

We choose a vector $\mathbf{k}\in\mathbb{R}^3$ with
\begin{equation}\label{choice of k}
\sqrt{1+ |\mathbf{k}|^2}= \lambda.
\end{equation}
Let $\chi\in C_0^\infty(\mathbb{R}^3)$ be a function with $\mathrm{supp}\chi\subset\big\{\y\in\mathbb{R}^3\big\arrowvert 1\leqslant |\y|\leqslant 2\big\},$\\  $\|\chi\|_{L_2(\mathbb{R}^3)}= 1$.
For $\mathbf{p}\in\mathbb{R}^3$ we define a family of operators $\Lambda_+(\mathbf{p})$ in $\mathbb{C}^4$ by
\begin{equation}\label{Lambda_+(k)}
\Lambda_+(\mathbf{p}):= \frac{1}{2}+ \frac{\al\cdot\mathbf{p}+ \beta}{2\sqrt{|\mathbf{p}|^2+ 1}}.
\end{equation}
Let $u(\mathbf{k})\in \mathrm{Ran}\big(\Lambda_+(\mathbf{k})\big)$ and $\big|u(\mathbf{k})\big|= 1$. Then \eqref{choice of k} implies
\begin{equation}\label{action on u}
(\al\cdot\mathbf{k}+ \beta)u(\mathbf{k})= \lambda u(\mathbf{k}).
\end{equation}
For a sequence $0< R_j\nearrow \infty, \; j\in\mathbb{N}$ we define
\begin{equation}\label{psi_j}
\psi_j(\y):= R_j^{-3/2}\chi(R_j^{-1}\y)e^{i\mathbf{k}\cdot\mathbf{y}}u(\mathbf{k}), \quad \y\in\mathbb{R}^{3}, \quad j\in\mathbb{N}.
\end{equation}
Assuming that 
\begin{equation}\label{R_j condition}
R_{j+ 1}\geqslant 2R_j,
\end{equation}
we get
\begin{equation}\label{orthonorm}
\langle\psi_j, \psi_k\rangle= \delta_{jk}, \quad j, k\in\mathbb{N}.
\end{equation}
\begin{lemma}\label{projector is not important}
For the sequence $\{\psi_j\}_{j= 1}^\infty$ we have
\begin{equation}\label{Lambda is not important}
\|\Lambda_+\psi_j- \psi_j\|_{L_2(\mathbb{R}^3, \mathbb{C}^4)}\underset{j\rightarrow\infty}{\longrightarrow} 0.
\end{equation}
\end{lemma}
\begin{proof}{}
Since the Fourier transform of $\psi_j$ is
\[
\hat\psi_j(\mathbf{p})= R_j^{3/2}\hat\chi\big(R_j(\mathbf{p}- \mathbf{k})\big)u(\mathbf{k}),
\]
one has
\begin{equation}\begin{split}\label{two terms}
\|&\Lambda_+\psi_j- \psi_j\|_{L_2(\mathbb{R}^3, \mathbb{C}^4)}\\ &\leqslant \Big\|\big(\Lambda_+(\mathbf{p})- 1\big)I_{B(R_j^{-1/2}, \mathbf{k})}(\mathbf{p})R_j^{3/2}\hat\chi\big(R_j(\mathbf{p}- \mathbf{k})\big)u(\mathbf{k})\Big\|_{L_2(\mathbb{R}^3, \mathbb{C}^4)}\\&+ \Big\|\big(\Lambda_+(\mathbf{p})- 1\big)I_{\mathbb{R}^3\setminus B(R_j^{-1/2}, \mathbf{k})}(\mathbf{p})R_j^{3/2}\hat\chi\big(R_j(\mathbf{p}- \mathbf{k})\big)u(\mathbf{k})\Big\|_{L_2(\mathbb{R}^3, \mathbb{C}^4)}.
\end{split}\end{equation}
Obviously
\[
\Big\|\big(\Lambda_+(\mathbf{p})- 1\big)- \big(\Lambda_+(\mathbf{k})- 1\big)\Big\|_{\mathbb{C}^4\rightarrow \mathbb{C}^4}\leqslant C|\mathbf{p}- \mathbf{k}|
\]
for $C> 0$ independent of $\mathbf{k}$. Hence one can estimate the first term in \eqref{two terms} by $CR_j^{-1/2}\|\chi\|_{L_2(\mathbb{R}^3)}$. The second term in \eqref{two terms} is bounded by $\|\hat\chi\|_{L_2(\mathbb{R}^3\setminus B(R_j^{1/2}))}$, and consequently converges to zero, too.
\end{proof}
\vspace{\baselineskip}

Now we are ready to define the desired Weyl sequence. Let
\begin{equation}\begin{split}\label{Psi}
\Psi_j(\x, \x_N):&= \varphi_j(\x)\otimes\psi_j(\x_N),\\ \x&= (\x_1, \dots, \x_{N- 1})\in\mathbb{R}^{3(N- 1)}, \quad \x_N\in\mathbb{R}^3, \quad j\in\mathbb{N}.
\end{split}\end{equation}
Recall that $\varphi_j$ is a vector function with $4^{(N- 1)}$ components and $\psi_j$ is a vector function with $4$ components.
\begin{lemma}\label{Lambda_+Psi_j}
The sequence $\{\Psi_j\}_{j= 1}^\infty$ has the following properties:\\
{\em (i)} $\quad\|\Lambda_+^N\Psi_j\|\underset{j\rightarrow\infty}{\longrightarrow}1$,\\
{\em (ii)} $\quad|\langle\Lambda_+^N\Psi_j, \Lambda_+^N\Psi_k\rangle|\rightarrow 0, \quad k\neq j, \quad k,j\rightarrow\infty$,
\begin{equation}\label{weyl decay}
\mathrm{(iii)} \quad\big\|(\mathcal{H}_N- E_{N- 1}- \lambda)\Lambda_+^N\Psi_j\big\|\underset{j\rightarrow\infty}{\longrightarrow}0.
\end{equation}
\end{lemma}
\begin{proof}{ of Lemma~\ref{Lambda_+Psi_j}}
Relations $(i)$ and $(ii)$ follow from Lemma~\ref{projector is not important} and relation \eqref{orthonorm}.\\
Let us prove \eqref{weyl decay}. We have
\begin{equation}\label{small}\begin{split}
&(\mathcal{H}_N- E_{N- 1}- \lambda)\Lambda_+^N\Psi_j\\&= (\mathcal{H}_{N- 1}- E_{N- 1})\Lambda_+^N\Psi_j+ \Lambda_+^N\bigg((D_N- \lambda)- \frac{\alpha Z}{|\x_N|}+ \underset{n= 1}{\overset{N- 1}{\sum}}\frac{\alpha}{|\x_n- \x_N|}\bigg)\Lambda_+^N\Psi_j.
\end{split}\end{equation}
The operator $\mathcal{H}_{N- 1}$ acts on the function $\varphi_j$ only.\\
The first term in \eqref{small} tends to zero in norm according to \eqref{varphi_j}.\\
Since $\Lambda_+^N$ commutes with $D_N$ we have
\begin{equation}\label{1st}\begin{split}
\big\|(D_N- \lambda)&\Lambda_+^N\Psi_j\big\|_{L_2(\mathbb{R}^{3N}, \mathbb{C}^{4^N})}\leqslant \big\|(D_N- \lambda)\Psi_j\big\|_{L_2(\mathbb{R}^{3N}, \mathbb{C}^{4^N})}\\ &\leqslant \big\|R_j^{-3/2}\al\cdot\nabla_{\mathbf{x}_N}\chi(R_j^{-1}\mathbf{x}_N)u(\mathbf{k})\big\|_{L_2(\mathbb{R}^{3}, \mathbb{C}^4)}\underset{j\rightarrow\infty}{\longrightarrow} 0.
\end{split}\end{equation}
In the second inequality of \eqref{1st} we have used \eqref{action on u}.\\
To prove \eqref{weyl decay} it suffices now to show that
\begin{equation}\label{potential terms}
\bigg\|\Lambda_+^N\bigg(-\frac{\alpha Z}{|\x_N|}+ \underset{n= 1}{\overset{N- 1}{\sum}}\frac{\alpha}{|\x_n- \x_N|}\bigg)\Lambda_+^N\Psi_j\bigg\|\underset{j\rightarrow\infty}{\longrightarrow}0.
\end{equation}
Let
\[
\eta_1\in C_0^\infty(\mathbb{R}^3), \quad \eta_1(\x)\equiv\begin{cases}1,& \x\in B(1/2),\\ 0,& \x\in\mathbb{R}^3\setminus B(3/4),\end{cases} \quad \eta_j(\x):= \eta_1(\x/R_j), \quad j\in\mathbb{N}.
\]

We estimate now the term in \eqref{potential terms} corresponding to the interaction with the nucleus.
\begin{equation}\begin{split}\label{2nd}
\Big\|\Lambda_+^N\frac{\alpha Z}{|\x_N|}\Lambda_+^N\Psi_j\Big\|&\leqslant \alpha Z\bigg(\Big\|\frac{\eta_j(\x_N)}{|\x_N|}(\Lambda_+\psi_j)(\x_N)\Big\|\\ &+ \Big\|\frac{\big(1- \eta_j(\x_N)\big)}{|\x_N|}(\Lambda_+\psi_j)(\x_N)\Big\|\bigg).
\end{split}\end{equation}
For the first term in \eqref{2nd} the Hardy inequality and the commutativity of $\Lambda_+$ with $\nabla$ imply
\begin{equation}\label{1stin2nd}
\Big\|\frac{\eta_j(\x_N)}{|\x_N|}(\Lambda_+\psi_j)(\x_N)\Big\|\leqslant 2\|\nabla\eta_j\|_{L_\infty(\mathbb{R}^3)}\|\Lambda_+\psi_j\|+ 2\|\Lambda_+\nabla\psi_j\|_{L_2(B(3R_j/4))}.
\end{equation}
The first term in \eqref{1stin2nd} decays as $R_j\rightarrow \infty$, since it is bounded by \\ $2\|\nabla\eta_1\|_{L_\infty(\mathbb{R}^3)}R_j^{-1}$. Applying Lemma~\ref{decay lemma} to the second term on the \rhs of \eqref{1stin2nd}, we arrive at
\begin{equation}\begin{split}\label{2ndin2nd}
\|\Lambda_+\nabla\psi_j&\|_{L_2(B(3R_j/4))}\leqslant \Big(\frac{4\pi}{3}\Big(\frac{3R_j}{4}\Big)^3\Big)^{1/2}\|\Lambda_+\nabla\psi_j\|_{L_\infty(B(3R_j/4))}\\ &\leqslant \Big(\frac{4\pi}{3}\Big(\frac{3R_j}{4}\Big)^3\Big)^{1/2}G(R_j/4)\Big(\frac{4\pi}{3}(2R_j)^3\Big)^{1/2}\|\nabla\psi_j\|_{L_2(\mathbb{R}^3)}.
\end{split}\end{equation}
The last factor in \eqref{2ndin2nd} is bounded uniformly in $j$. The function $G$ and consequently the \rhs of \eqref{2ndin2nd} decays exponentially as $R_j\rightarrow\infty$. We can estimate the second term in \eqref{2nd} by $2\alpha Z/R_j$. Hence the \rhs of \eqref{2nd} tends to $0$ as $R_j\rightarrow\infty$.

Let us turn to the interaction between the $N^{th}$ and the $n^{th}$ electrons. We have
\begin{equation}\label{3rd}\begin{split}
\Big\|\Lambda_+^N\frac{\alpha}{|\x_n- \x_N|}\Lambda_+^N\Psi_j\Big\|&\leqslant 
\Big\|\frac{\alpha\rho_{\varphi_j}(\x_n)}{|\x_n- \x_N|}(\Lambda_+\psi_j)(\x_N)\Big\|_{\{|\x_n|\geqslant R_j/4\}}\\&+ \Big\|\frac{\alpha\rho_{\varphi_j}(\x_n)}{|\x_n- \x_N|}\eta_j(\x_N)(\Lambda_+\psi_j)(\x_N)\Big\|_{\{|\x_n|< R_j/4\}}\\&+ \Big\|\frac{\alpha\rho_{\varphi_j}(\x_n)}{|\x_n- \x_N|}\big(1- \eta_j(\x_N)\big)(\Lambda_+\psi_j)(\x_N)\Big\|_{\{|\x_n|< R_j/4\}},
\end{split}\end{equation}
where
\begin{equation}\label{rho}
\rho_{\varphi_j}(\x_n)= \int\limits_{\mathbb{R}^{3(N- 2)}}\big|\varphi_j(\x_1,\dots,\x_{N- 1})\big|^2d\x_1\cdots d\x_{n- 1}d\x_{n+ 1}\cdots d\x_{N- 1}
\end{equation}
if $N> 2$, and $\rho_{\varphi_j}= |\varphi_j|^2$ if $N= 2$; $\|\rho_{\varphi_j}\|_{L_2(\mathbb{R}^3)}= 1$. By the Hardy inequality the fist term in \eqref{3rd} can be estimated as
\begin{equation}\label{1stin3rd}
\Big\|\frac{\alpha\rho_{\varphi_j}(\x_n)}{|\x_n- \x_N|}(\Lambda_+\psi_j)(\x_N)\Big\|_{\{|\x_n|\geqslant R_j/4\}}\leqslant 2\alpha\|\nabla\psi_j\|\|\rho_{\varphi_j}\|_{L_2(\mathbb{R}^3\setminus B(R_j/4))}.
\end{equation}
For the second term in \eqref{3rd} analogously to \eqref{1stin2nd} and \eqref{2ndin2nd} we have
\begin{equation}\begin{split}\label{2ndin3rd}
\Big\|\frac{\alpha\rho_{\varphi_j}(\x_n)}{|\x_n- \x_N|}\eta_j(\x_N)(\Lambda_+\psi_j)(\x_N)\Big\|_{\{|\x_n|< R_j/4\}}\leqslant 2\alpha\|\nabla\eta_1\|_{L_\infty(\mathbb{R}^3)}R_j^{-1}\\ + 2\alpha\Big(\frac{4\pi}{3}\Big(\frac{3R_j}{4}\Big)^3\Big)^{1/2}G(R_j/4)\Big(\frac{4\pi}{3}(2R_j)^3\Big)^{1/2}\|\nabla\psi_j\|_{L_2(\mathbb{R}^3)}.
\end{split}\end{equation}
Finally, the last term on the \rhs of \eqref{3rd} can be estimated by $4\alpha/R_j$, since on the domain of integration $|\x_n- \x_N|> R_j/4$.

Combining these estimates, we arrive at \eqref{weyl decay}. Lemma~\ref{Lambda_+Psi_j} is proved.
\end{proof}

Our next goal is to prove that the functions $\{\Psi_j\}_{j= 1}^\infty$ can be antisymmetrized without violation of the condition \eqref{lower norm bound} at least for $j$ big enough.

Let 
\begin{equation}\begin{split}\label{T_kN}
(T_{kN}\Psi_j)(\x_1, \dots, \x_N):= \varphi(\x_1, \dots, \x_{k- 1}, \x_N, \x_{k+ 1}, \dots, \x_{N- 1})\otimes\psi_j(\x_k),\\ \quad k=1, \dots, N- 1, \quad T_{NN}\Psi_j:= -\Psi_j, \quad j\in\mathbb{N}.
\end{split}\end{equation}
The operator $T_{kN}$ permutes the $k^{th}$ and the $N^{th}$ electrons.
The functions $P_A\Psi_j$ are given by
\begin{equation}\label{Phi}
(P_A\Psi_j)(\x_1,\dots, \x_N)= \frac{-1}{\sqrt{N}}\underset{k= 1}{\overset{N- 1}{\sum}}T_{kN}\Lambda_+^N\Psi_j(\x_1,\dots, \x_N), \quad j\in\mathbb{N}.
\end{equation}
For the norms of these functions we have
\begin{equation}\begin{split}\label{normalization}
\|P_A\Psi_j\|^2= \frac{1}{N}\langle\underset{k= 1}{\overset{N- 1}{\sum}}T_{kN}\varphi_j(\x_1,\dots, \x_{N- 1})&(\Lambda_+\psi_j)(\x_N),\\ &\underset{l= 1}{\overset{N- 1}{\sum}}T_{lN}\varphi_j(\x_1,\dots, \x_{N- 1})(\Lambda_+\psi_j)(\x_N)\rangle \\ \geqslant 1- \frac{N-1}{2}\big|\langle\varphi_j(\x_1, \dots, \x_{N- 1})(\Lambda_+&\psi_j)(\x_N), (\Lambda_+\psi_j)(\x_1)\varphi_j(\x_2, \dots, \x_N)\rangle\big|,
\end{split}\end{equation}
where the scalar products are taken in $\underset{n= 1}{\overset{N}{\otimes}}L_2(\mathbb{R}^3, \mathbb{C}^4)$.
To estimate the inner product on the \rhs of \eqref{normalization}, we write
\begin{equation}\begin{split}\label{inner product estimate}
\big|\langle\varphi_j(\x_1, \dots, \x_{N- 1})(\Lambda_+\psi_j)(\x_N), (\Lambda_+\psi_j)(\x_1)\varphi_j&(\x_2, \dots, \x_N)\rangle\big|\\ \leqslant \big|\langle \big(1- I_{B(R_j/2)}(\x_1)\big)\varphi_j(\x_1, \dots, \x_{N- 1})(\Lambda_+\psi_j)(\x_N),&\\ (\Lambda_+\psi_j)&(\x_1)\varphi_j(\x_2, \dots, \x_N)\rangle\big|\\ + \Big|\langle\varphi_j(\x_1, \dots, \x_{N- 1})(\Lambda_+\psi_j)(\x_N), I_{B(R_j/2)}(\x_1)(\Lambda_+\psi_j)&(\x_1)\varphi_j(\x_2, \dots, \x_N)\rangle\Big|.
\end{split}\end{equation}
The first term on the \rhs of \eqref{inner product estimate} tends to zero since\\ $\Big\|\big(1- I_{B(R_j/2)}(\x_1)\big)\varphi_j(\x_1, \dots, \x_{N- 1})\Big\|_{L_2(\mathbb{R}^{N- 1}, \mathbb{C}^{4^{(N- 1)}})}$ does. The second one also vanishes, because by Lemma~\ref{decay lemma}
\[
\Big\|I_{B(R_j/2)}(\x_1)(\Lambda_+\psi_j)(\x_1)\Big\|\underset{R_j\rightarrow \infty}{\longrightarrow}0.
\]
This completes the proof of \eqref{easy claim}.

\section{Proof of Theorem~\ref{HVZ theorem}: ``Hard Part''}\label{hard part}

We shall prove that 
\begin{equation}\label{hard claim}
\inf~\sigma_{\mathrm{ess}}(\mathcal{H}_N)\geqslant E_{N-1}+ 1.
\end{equation}
Together with \eqref{easy claim} this gives \eqref{HVZ}.

\subsection{Partition of Unity}

For a function $\chi\in C^\infty\big(\mathbb{R}^3, [0, 1]\big)$ with
$\chi\arrowvert_{B(1)}\equiv 0$ and
$\chi\arrowvert_{\mathbb{R}^3\setminus B(2)}\equiv 1$, let
\[
\tilde\chi_a:= \chi(\x_a), \quad a=1\dots N, \quad
\tilde\chi_0:=\underset{a=1}{\overset{N}{\prod}}(1- \tilde\chi_a).
\]
Let
\[
\varphi(\x):=\underset{a= 0}{\overset{N}{\sum}}\tilde\chi_a^2(\x).
\]
Obviously, there exists a constant $\delta> 0$ such that for any $\x\in\mathbb{R}^{3N}$ we have $\delta< \varphi(\x)< \delta^{-1}$.
For $R>0$ we define a partition of unity
\begin{equation}\label{partition}
\chi_a(\x):=\frac{\tilde\chi_a(\x/R)}{\sqrt{\varphi(\x/R)}}, \quad
\x\in\mathbb{R}^{3N}, \quad a= 0, \dots, N.
\end{equation}
It is clear that the partition \eqref{partition} satisfies all the hypotheses \eqref{partition requirements} of Lemma~\ref{localization lemma}. Moreover, for
$a\neq 0$ the functions $\chi_a$ are symmetric under all permutations of the electrons which do not include the $a^{th}$ one. We also note that the derivatives of $\chi_a$ decay as $R$ tends to infinity:
\begin{equation}\label{derivatives decay}
\|\nabla\chi_a\|_\infty\leqslant CR^{-1}, \quad
\|\partial^2\chi_a\|_\infty\leqslant CR^{-2}, \quad a=0, \dots, N.
\end{equation}

\subsection{Estimates Outside the Compact Region}

For $\varepsilon> 0$ we choose $R= R(\varepsilon)$ big enough so that the following conditions hold:\\
(i)\\
\begin{equation}\label{condition1}
\widetilde C_{N, Z}\sum_{a\in\mathcal{A}}\big(\|\nabla\chi_a\|_{L_\infty(\mathbb{R}^{3N})}+ \|\partial^2\chi_a\|_{L_\infty(\mathbb{R}^{3N})}\big)\big(1+ \|\nabla\chi_a\|_{L_\infty(\mathbb{R}^{3N})}\big)^2< \varepsilon,
\end{equation}
where $\widetilde C_{N, Z}$ is the constant in \eqref{localization formula},\\
(ii)\\
\begin{equation}\label{condition2}
\frac{\alpha Z}{R}< \varepsilon(1- \alpha Z),
\end{equation}
(iii)\quad For any $\psi\in\Lambda^N_+\underset{n= 1}{\overset{N}{\wedge}}H^{1/2}(\mathbb{R}^3, \mathbb{C}^4)$ and $a= 1,\dots, N$
\begin{equation}\label{condition3}\begin{split}
\bigg|\langle\Big(E_{N- 1}+ 1&- \frac{\alpha Z}{|\x_a|}\Big)[\Lambda^N_+, \chi_a]\psi, \Lambda_+^N\chi_a\psi\rangle\\ &+ \langle\Big(E_{N- 1}+ 1- \frac{\alpha Z}{|\x_a|}\Big)\chi_a\psi, [\Lambda_+^N, \chi_a]\psi\rangle\bigg|< \frac{\varepsilon}{N}\langle\mathcal{H}_N\psi, \psi\rangle.
\end{split}\end{equation}
The possibility to fulfil \eqref{condition3} choosing $R$ big enough follows from the Kato inequality and Lemmata~\ref{multiplicator lemma}, \ref{boundedness of many-particle commutator in H^1/2}, and \ref{control lemma}.

We now estimate from below the quadratic form of $\mathcal{H}_N$ on a function $\psi$ from $\Lambda_+^N\underset{n= 1}{\overset{N}{\wedge}}H^{1/2}(\mathbb{R}^3, \mathbb{C}^4)$. Recall that $\Lambda_+^N\underset{n= 1}{\overset{N}{\wedge}}H^{1/2}(\mathbb{R}^3, \mathbb{C}^4)$ is the form domain of $\mathcal{H}_N$.
By Lemma~\ref{localization lemma} and \eqref{condition1}
\begin{equation}\label{cutted}
\langle\mathcal{H}_N\psi, \psi\rangle\geqslant \underset{a= 0}{\overset{N}{\sum}}\langle\mathcal{H}_N\Lambda_+^N\chi_a\psi, \Lambda_+^N\chi_a\psi\rangle- \varepsilon\langle\mathcal{H}_N\psi, \psi\rangle.
\end{equation}
For $a= 1, \dots, N$ we have
\begin{equation}\label{detalized}
\langle\mathcal{H}_N\Lambda_+^N\chi_a\psi, \Lambda_+^N\chi_a\psi\rangle= \langle\Big(\mathcal{H}_{N-1}+ D_a- \frac{\alpha Z}{|\x_a|}+ \underset{k\neq a}{\sum}\frac{\alpha}{|\x_a- \x_k|}\Big)\Lambda^N_+\chi_a\psi, \Lambda_+^N\chi_a\psi\rangle,
\end{equation}
where $\mathcal{H}_{N- 1}$ acts on the coordinates of all electrons except the $a^{th}$ one. The inequalities $\mathcal{H}_{N- 1}\geqslant E_{N- 1},\: \Lambda_+^ND_a\Lambda_+^N\geqslant 1$, and $\underset{k\neq a}{\sum}\frac{\D\alpha}{\D|\x_a- \x_k|}> 0$ for $a= 1, \dots, N$ imply
\begin{equation}\begin{split}\label{commutations}
\langle\mathcal{H}_N\Lambda_+^N\chi_a\psi, \Lambda_+^N\chi_a\psi\rangle&\geqslant \langle\Big(E_{N- 1}+ 1- \frac{\alpha Z}{|\x_a|}\Big)\Lambda^N_+\chi_a\psi, \Lambda_+^N\chi_a\psi\rangle\\ = \langle\Big(E_{N- 1}+ 1- \frac{\alpha Z}{|\x_a|}\Big)\chi_a\psi, \chi_a\psi\rangle&+ \langle\Big(E_{N- 1}+ 1- \frac{\alpha Z}{|\x_a|}\Big)[\Lambda^N_+, \chi_a]\psi, \Lambda_+^N\chi_a\psi\rangle\\ &+ \langle\Big(E_{N- 1}+ 1- \frac{\alpha Z}{|\x_a|}\Big)\chi_a\psi, [\Lambda_+^N, \chi_a]\psi\rangle.
\end{split}\end{equation}
Since on $\mathrm{supp}~\chi_a$ we have $|\x_a|\geqslant R$, from the relations \eqref{condition2}, \eqref{condition3}, and \eqref{commutations} we conclude
\begin{equation}\label{each term}\begin{split}
\langle\mathcal{H}_N\Lambda_+^N\chi_a\psi, \Lambda_+^N\chi_a\psi\rangle&\geqslant (E_{N- 1}+ 1)\langle\chi_a\psi, \chi_a\psi\rangle\\ &- \frac{\varepsilon}{N}\langle\mathcal{H}_N\psi, \psi\rangle- \varepsilon(1- \alpha Z)\|\psi\|^2, \quad a= 1, \dots, N.
\end{split}\end{equation}
Using \eqref{semiboundedness}, we arrive at
\begin{equation}\label{N-semiboundedness}
\langle\mathcal{H}_N\psi, \psi\rangle\geqslant \langle\underset{n= 1}{\overset{N}{\sum}}\Big(D_n- \frac{\alpha Z}{|\x_n|}\Big)\psi, \psi\rangle\geqslant N(1- \alpha Z)\|\psi\|^2.
\end{equation}
Due to \eqref{cutted}, \eqref{each term}, and \eqref{N-semiboundedness}
\begin{equation}\label{halfway}
(1+ 3\varepsilon)\langle\mathcal{H}_N\psi, \psi\rangle\geqslant (E_{N- 1}+ 1)\underset{a= 1}{\overset{N}{\sum}}\langle\chi_a\psi, \chi_a\psi\rangle+ \langle\mathcal{H}_N\Lambda_+^N\chi_0\psi, \Lambda_+^N\chi_0\psi\rangle.
\end{equation}

\subsection{Estimate Inside the Compact Region}

Our next goal is to estimate from below the quadratic form of the operator $H_N\Lambda_+^N$ on the function $\chi_0\psi$ supported in $[-2R, 2R]^{3N}$. 

\begin{lemma}\label{lemma2}
For $M> 0$ let $W_M:=\big\{\mathbf{p}\in\mathbb{R}^{3N}\big\arrowvert |p_i|\leqslant M, i=1,\dots,3N\big\}, \:\widetilde W_M:=\mathbb{R}^{3N}\setminus W_M$. There exists a finite set $Q_M$ of functions in $L_2(\mathbb{R}^{3N})$ such that for any function $f\in L_2(\mathbb{R}^{3N})$ with $\mathrm{supp}f\subset[-2R, 2R]^{3N},\: f\bot Q_M$ holds
\begin{equation}\label{much is away}
\|\hat f\|_{L_2(\widetilde W_M)}\geqslant \frac{1}{2}\|\hat f\|_{L_2(\mathbb{R}^{3N})}. 
\end{equation}
\end{lemma}

The proof of Lemma~\ref{lemma2} is analogous to the proof of Theorem~7 in \cite{VugalterWeidl2003} and will be given in the Appendix~\ref{proof of the W lemma} for convenience.

\medskip

It follows from \eqref{without Coulomb} that for any $M> 0$
\begin{equation}\begin{split}\label{step1}
\langle \mathcal{H}_N\Lambda_+^N\chi_0\psi, \Lambda_+^N\chi_0\psi\rangle&= \langle\bigg(\underset{n= 1}{\overset{N}{\sum}}\Big(D_n- \frac{\alpha Z}{|\x_n|}\Big)+ \underset{n< j}{\sum}\frac{\alpha}{|\x_n- \x_j|}\bigg)\Lambda_+^N\chi_0\psi, \Lambda_+^N\chi_0\psi\rangle\\ &\geqslant \frac{Z_c- Z}{Z_c}\langle\underset{n= 1}{\overset{N}{\sum}}D_n I_{\widetilde W_M}\Lambda_+^N\chi_0\psi, \Lambda_+^N\chi_0\psi\rangle.
\end{split}\end{equation}
Here $I_{\widetilde W_M}$ is the operator of multiplication by the characteristic function of $\widetilde W_M$ in momentum space.

We choose
\begin{equation}\label{M}
M:= \sqrt{\Big(\frac{8Z_c(E_{N- 1}+ 1)}{Z_c- Z}\Big)^2- 1} 	
\end{equation}
and assume henceforth that $f:= \chi_0\psi$ is orthogonal to the set $Q_M$ defined in Lemma~\ref{lemma2}. Since in momentum space the operator $D$ acts on functions from $\Lambda_+L_2(\mathbb{R}^3, \mathbb{C}^4)$ as multiplication by $\sqrt{|\mathbf{k}|^2+ 1}$, we have
\begin{equation}\label{big kinetic energy}
\langle\underset{n= 1}{\overset{N}{\sum}}D_n I_{\widetilde W_M}\Lambda_+^N\chi_0\psi, \Lambda_+^N\chi_0\psi\rangle\geqslant \sqrt{M^2+ 1}\|I_{\widetilde W_M}\Lambda_+^N\chi_0\psi\|^2.
\end{equation}
Inequalities \eqref{step1} and \eqref{big kinetic energy} imply
\begin{equation}\begin{split}\label{step2}
\langle\mathcal{H}_N\Lambda_+^N\chi_0\psi&, \Lambda_+^N\chi_0\psi\rangle\geqslant \frac{Z_c- Z}{Z_c}\sqrt{M^2+ 1}\|I_{\widetilde W_M}\Lambda_+^N\chi_0\psi\|^2\\ \geqslant \frac{Z_c- Z}{Z_c}&\sqrt{M^2+ 1}\Big(\|I_{\widetilde W_M}\chi_0\psi\|- \big\|I_{\widetilde W_M}[\Lambda_+^N, \chi_0]\psi\big\|\Big)^2\\ \geqslant \frac{Z_c- Z}{Z_c}&\sqrt{M^2+ 1}\Big(\frac{1}{2}\|I_{\widetilde W_M}\chi_0\psi\|^2- \big\|I_{\widetilde W_M}[\Lambda_+^N, \chi_0]\psi\big\|^2\Big)\\ \geqslant 4(E_{N- 1}&+ 1)\|I_{\widetilde W_M}\chi_0\psi\|^2\\&- 8(E_{N- 1}+ 1)\big\|[\Lambda_+^N, \chi_0]\big\|_{L_2(\mathbb{R}^{3N}, \mathbb{C}^{4^N})\rightarrow L_2(\mathbb{R}^{3N}, \mathbb{C}^{4^N})}^2\|\psi\|^2.
\end{split}\end{equation}
In the last step we used our choice of $M$ (see \eqref{M}). Since $\Lambda_+$ is a bounded operator in $L_2(\mathbb{R}^3, \mathbb{C}^4)$, and
\[
\big\|[\chi_0, \Lambda_+^N]\big\|_{L_2(\mathbb{R}^{3N}, \mathbb{C}^{4^N})\rightarrow
L_2(\mathbb{R}^{3N}, \mathbb{C}^{4^N})}\leqslant C\big(\|\nabla\chi_0\|_\infty+ \|\partial^2\chi_0\|_\infty\big)
\]
by Lemma~\ref{commutator lemma} and the relation \eqref{reduction to one particle}, we can choose $R$ big enough, so that 
\begin{equation}\label{again varepsilon}
8|E_{N- 1}+ 1|\big\|[\Lambda_+^N, \chi_0]\big\|_{L_2(\mathbb{R}^{3N}, \mathbb{C}^{4^N})\rightarrow L_2(\mathbb{R}^{3N}, \mathbb{C}^{4^N})}^2\leqslant \varepsilon.
\end{equation}
For the first term on the \rhs of \eqref{step2} Lemma~\ref{lemma2} implies
\begin{equation}\label{big away}
4\|I_{\widetilde W_M}\chi_0\psi\|^2\geqslant \|\chi_0\psi\|^2.
\end{equation}
As a consequence of \eqref{step2} --- \eqref{big away}, we have
\begin{equation}\label{another half}
\langle\mathcal{H}_N\Lambda_+^N\chi_0\psi, \Lambda_+^N\chi_0\psi\rangle\geqslant (E_{N- 1}+ 1)\|\chi_0\psi\|^2- \varepsilon\|\psi\|^2.
\end{equation}

\subsection{Completion of the Proof}\label{completion}

By \eqref{halfway} and \eqref{another half}
\begin{equation}\label{final}
(1+ 3\varepsilon)\langle \mathcal{H}_N\psi, \psi\rangle\geqslant (E_{N- 1}+ 1)\underset{a= 0}{\overset{N}{\sum}}\langle\chi_a\psi,
  \chi_a\psi\rangle- \varepsilon\|\psi\|^2= (E_{N- 1}+ 1- \varepsilon)\|\psi\|^2
\end{equation}
for any $\varepsilon> 0$ and any $\psi$ in the form domain of $\mathcal{H}_N$ orthogonal to the finite set of functions (cardinality of this set depends on $\varepsilon$). This implies the discreteness of the spectrum of $\mathcal{H}_N$ below $E_{N- 1}+ 1$ and thus \eqref{hard claim}.

\section{Existence of Eigenvalues}\label{existence of eigenvalues}

\paragraph{1.}

To prove the infiniteness of the discrete spectrum of $\mathcal{H}_N$ it suffices to construct for a given $Q\in\mathbb{N}$ a $Q$--dimensional subspace $\mathcal{M}$ such that for any $\Psi\in\mathcal{M}$ we have $\langle\mathcal{H}_N\Psi, \Psi\rangle< (E_{N- 1}+ 1)\|\Psi\|^2$. 

Using induction on $N$ and the well--known existence of the ground state of $\mathcal{H}_1$, we can assume that $\mathcal{H}_{N-1}$ has a ground state $\phi$. Let $\widetilde\psi\in C_0^\infty(\mathbb{R}^3, \mathbb{C}^4)$ be a function with $\mathrm{supp~}\widetilde\psi\subset B(N- \frac{1}{5})\setminus B(N- \frac{2}{5})$,  whose $3^{rd}$ and $4^{th}$ components are identical to zero. Let $\|\widetilde\psi\|_{L_2(\mathbb{R}^3, \mathbb{C}^4)}= 1$. Let
\begin{equation}\label{psi}
\psi_m(\y):= R_m^{-3/2}\widetilde\psi\Big(\frac{\D\y}{\D R_m}\Big), \quad R_m:= 2^mR, \quad m=1, \dots Q.
\end{equation}
The parameter $R$ will be chosen later. Note that $\Lambda_+\psi_m\neq 0$ for large $R$ due to \eqref{Lambda_+} and the choice of the components of $\widetilde\psi$.

We consider the quadratic form of $\mathcal{H}_N$ on linear combinations of the form
\begin{equation}\label{test functions}
\underset{m= 1}{\overset{Q}{\sum}}c_m\underset{k= 1}{\overset{N}{\sum}}T_{kN}(\phi\otimes\Lambda_+\psi_m).
\end{equation}
Here $T_{kN}$ for $k< N$ is the operator permuting the $k^{th}$ and the $N^{th}$ electrons, $T_{NN}:= -1$. In the tensor product $\phi\otimes\Lambda_+\psi_m$ the function $\phi$ is assumed to depend on $\x_1,\dots, \x_{N- 1}$, and $\psi_m$ depends on $\x_N$. The functions \eqref{test functions} are antisymmetric in all variables.

It suffices to show that on the functions \eqref{test functions} the quadratic form of
\[
\widetilde{\mathcal{H}}_N:= \mathcal{H}_N- E_{N- 1}- 1
\]
is negative for any choice of the coefficients $\{c_m\}_{m= 1}^Q$.

Using the permutation symmetry of $\phi$ and $\widetilde{\mathcal{H}}_N$, we can write
\begin{equation}\begin{split}\label{expansion}
\langle\widetilde{\mathcal{H}}_N\underset{m= 1}{\overset{Q}{\sum}}c_m\underset{k= 1}{\overset{N}{\sum}}T_{kN}(\phi\otimes\Lambda_+\psi_m), \underset{n= 1}{\overset{Q}{\sum}}c_n\underset{l= 1}{\overset{N}{\sum}}T_{lN}(\phi\otimes\Lambda_+\psi_n)\rangle\\ = \underset{m= 1}{\overset{Q}{\sum}}|c_m|^2\underset{k, l= 1}{\overset{N}{\sum}}\langle\widetilde{\mathcal{H}}_N T_{kN}(\phi\otimes\Lambda_+\psi_m), T_{lN}(\phi\otimes\Lambda_+\psi_m)\rangle\\ + 2\underset{n< m}{\sum}c_m\overline{c_n}\underset{k, l= 1}{\overset{N}{\sum}}\langle\widetilde{\mathcal{H}}_N T_{kN}(\phi\otimes\Lambda_+\psi_m), T_{lN}(\phi\otimes\Lambda_+\psi_n)\rangle\\ \leqslant \underset{m= 1}{\overset{Q}{\sum}}|c_m|^2\bigg\{N\langle \widetilde{\mathcal{H}}_N (\phi\otimes\Lambda_+\psi_m), \phi\otimes\Lambda_+\psi_m\rangle\\ + \frac{N(N- 1)}{2}\underset{n= 1}{\overset{Q}{\sum}}\big|\langle\widetilde{\mathcal{H}}_N (\phi\otimes\Lambda_+\psi_m), T_{1N}(\phi\otimes\Lambda_+\psi_n)\rangle\big|\\ + N\underset{n\neq m}{\overset{Q}{\sum}}\big|\langle\widetilde{\mathcal{H}}_N (\phi\otimes\Lambda_+\psi_m), \phi\otimes\Lambda_+\psi_n\rangle\big|\bigg\}.
\end{split}\end{equation}
Our strategy is to show that the first term on the \rhs of \eqref{expansion} is negative and of the order $R^{-1}$ as $R\rightarrow \infty$, whereas the other terms decay more rapidly.

\paragraph{2.}

For the first term on the \rhs of \eqref{expansion} we have
\begin{equation}\begin{split}\label{on trial}
&\langle\widetilde{\mathcal{H}}_N(\phi\otimes\Lambda_+\psi_m), \phi\otimes\Lambda_+\psi_m\rangle\\& = \langle (D- 1)\Lambda_+\psi_m, \Lambda_+\psi_m\rangle- \langle\frac{\alpha Z}{|\x|}\Lambda_+\psi_m, \Lambda_+\psi_m\rangle\\ &+ \underset{i< N}{\sum}\langle\frac{\alpha}{|\x_i- \x_N|}\rho_\phi(\x_i)(\Lambda_+\psi_m)(\x_N), \rho_\phi(\x_i)(\Lambda_+\psi_m)(\x_N)\rangle
\end{split}\end{equation}
with $\rho_\phi$ defined in \eqref{rho}. 

We start with the lower bound on the first term on the \rhs of \eqref{on trial}.
Recall that $\psi_1$ has the last two components equal to zero, thus relations \eqref{psi}, \eqref{Dirac}, and \eqref{Lambda_+} imply
\begin{equation}\begin{split}\label{kinetic}
\langle(D\Lambda_+&- \Lambda_+)\psi_m, \psi_m\rangle\\= &\langle\Big(\frac{1}{2}+ \frac{\sqrt{|\mathbf{q}|^2+ R_m^2}}{2R_m}- \frac{1}{2}- \frac{R_m}{2\sqrt{|\mathbf{q}|^2+ R_m^2}}\Big)\hat\psi_1(\mathbf{q}), \hat\psi_1(\mathbf{q})\rangle\\&= \langle\frac{|\mathbf{q}|^2}{2R_m\sqrt{|\mathbf{q}|^2+ R_m^2}}\hat\psi_1(\mathbf{q}), \hat\psi_1(\mathbf{q})\rangle\leqslant \frac{1}{2R_m^2}\big\||\mathbf{q}|\hat\psi_1(\mathbf{q})\big\|^2.
\end{split}\end{equation}
Here $\mathbf{q}$ is the momentum dual to $\y$ in \eqref{psi}. The norm in \eqref{kinetic} is finite, since $\tilde\psi_1\in C_0^\infty(\mathbb{R}^3, \mathbb{C}^4)$. 

To estimate the interaction of an electron with the nucleus we choose
\begin{equation}\label{delta}
\delta:= \frac{1}{10N- 8}.
\end{equation}
Obviously,
\begin{equation}\begin{split}\label{ii}
&-\langle\frac{\alpha Z}{|\x|}\Lambda_+\psi_m, \Lambda_+\psi_m\rangle\\&= -\langle\frac{\alpha Z}{|\x|}I_{B(ZR_m)}\Lambda_+\psi_m, \Lambda_+\psi_m\rangle -\langle\frac{\alpha Z}{|\x|}(1- I_{B(ZR_m)})\Lambda_+\psi_m, \Lambda_+\psi_m\rangle\\ &\leqslant -\frac{\alpha}{R_m}\Big(\|\Lambda_+\psi_m\|- \big\|(1- I_{B(ZR_m)})\Lambda_+\psi_m\big\|\Big)^2+ \frac{\alpha}{R_m}\big\|(1- I_{B(ZR_m)})\Lambda_+\psi_m\big\|^2\\ &\leqslant -\frac{(1- \delta)\alpha}{R_m}\|\Lambda_+\psi_m\|^2+ \frac{(2+ \delta^{-1})\alpha}{R_m}\big\|(1- I_{B(ZR_m)})\Lambda_+\psi_m\big\|^2. \end{split}\end{equation}
Lemma \ref{decay lemma} implies
\[\begin{split}
\big\|(1- I_{B(ZR_m)})\Lambda_+\psi_m\big\|^2\\ \leqslant \int\limits_{ZR_m}^\infty &4\pi r^2G^2\Big(r- \big(N- \frac{1}{5}\big)R_m\Big)\frac{4\pi}{3}\Big(N- \frac{1}{5}\Big)^3R_m^3\|\psi_m\|^2dr\\ &\leqslant \frac{16\pi^2}{3}\Big(N- \frac{1}{5}\Big)^3R_m^3\int\limits_{ZR_m}^\infty G^2\Big(r- \big(N- \frac{1}{5}\big)R_m\Big)r^2dr,
\end{split}\]
which is a function decaying faster than any power of $R_m$. 

Now we turn to estimating the electron--electron interaction. We split the corresponding quadratic form into three integrals:
\begin{equation}\begin{split}\label{iii}
\langle\frac{\alpha}{|\x_i- \x_N|}\rho_\phi(\x_i)(\Lambda_+\psi_m)(\x_N), \rho_\phi(\x_i)(\Lambda_+\psi_m)(\x_N)\rangle&\\= \alpha\bigg(\iint\limits_{|\x|\geqslant R_m/5}+ \iint\limits_{\substack{|\x|< R_m/5\\ |\y|< (N- 3/5)R_m}}+ \iint\limits_{\substack{|\x|< R_m/5\\ |\y|\geqslant (N- 3/5)R_m}}\bigg)\frac{\big|\rho_\phi(\x)\big|^2}{|\x- \y|}&\big|(\Lambda_+\psi_m)(\y)\big|^2d\x d\y\\&=: I_1+ I_2+ I_3.
\end{split}\end{equation}
From the Kato inequality (see \cite{Kato1966}, inequality V.5.33; see also \cite{Herbst1977})
\begin{equation}\label{Kato}
\int\limits_{\mathbb{R}^3}|\x|^{-1}\big|u(\x)\big|^2d\x\leqslant \frac{\pi}{2}\int\limits_{\mathbb{R}^3}|\mathbf{k}|\big|\hat u(\mathbf{k})\big|^2d\mathbf{k}, \quad \forall u\in H^{1/2}(\mathbb{R}^3)
\end{equation}
it follows that
\begin{equation}\label{I_1}
I_1\leqslant \frac{\pi\alpha}{2R_m}\|\rho_\phi\|^2_{L_2(\mathbb{R}^3\setminus B(R_m/5))}\big\||\mathbf{q}|^{1/2}\hat\psi_1(\mathbf{q})\big\|_{L_2(\mathbb{R}^3)}^2.
\end{equation}
Using Lemma \ref{decay lemma}, we arrive at
\begin{equation}\label{I_2}
I_2\leqslant \frac{4\pi\alpha}{3}\Big(N- \frac{1}{5}\Big)^3R_m^3G^2\Big(\frac{R_m}{5}\Big)\|\Lambda_+\psi_m\|^2\hspace{-0.4cm}\int\limits_{|\x|< \frac{R_m}{5}}\hspace{-0.3cm}\big|\rho_\phi(\x)\big|^2\hspace{-0.5cm}\int\limits_{|\y|< (N- \frac{3}{5})R_m}\hspace{-0.3cm}\frac{1}{|\x- \y|}d\y d\x.
\end{equation}
Here
\[
\int\limits_{|\y|< (N- 3/5)R_m}\frac{1}{|\x- \y|}d\y\leqslant \int\limits_{|\y|< (N- 3/5)R_m}\frac{1}{|\y|}d\y\leqslant 2\pi\Big(N- \frac{3}{5}\Big)^2R_m^2.
\]
Obviously, the \rhs$\!\!$'s of \eqref{I_1} and \eqref{I_2} decay faster than $R_m^{-1}$.
Finally, for $I_3$ we have
\begin{equation}\label{I_3}
I_3\leqslant \frac{\alpha}{\big(N- \frac{4}{5}\big)R_m}\|\Lambda_+\psi_m\|^2.
\end{equation}
Combining \eqref{on trial} --- \eqref{I_3}, for large $R$ we get
\begin{equation}\begin{split}\label{existence}
\langle\widetilde{\mathcal{H}}_N(\phi\otimes\Lambda_+\psi_m)&, \phi\otimes\Lambda_+\psi_m\rangle\\ &\leqslant \alpha\bigg(\frac{N- 1}{\big(N- \frac{4}{5}\big)}- 1+ \delta\bigg)\|\Lambda_+\psi_m\|^2R_m^{-1}+ o(R^{-1}).
\end{split}\end{equation}
The coefficient at $R_m^{-1}$ is negative due to \eqref{delta}.

\paragraph{3.}

Let us prove now that the second term on the \rhs of \eqref{expansion} decays faster than $R^{-1}$.
\begin{equation}\begin{split}\label{second term}
\langle\widetilde{\mathcal{H}}_N (\phi\otimes&\Lambda_+\psi_m), T_{1N}(\phi\otimes\Lambda_+\psi_n)\rangle\\ =
\langle\bigg((&\mathcal{H}_{N- 1}- E_{N- 1})+ (D_N- 1)- \frac{\alpha Z}{|\x_N|}+ \underset{j= 1}{\overset{N- 1}{\sum}}\frac{\alpha}{|\x_j- \x_N|}\bigg)\\ &\times \phi(\x_1,\dots, \x_{N- 1})(\Lambda_+\psi_m)(\x_N), (\Lambda_+\psi_n)(\x_1)\phi(\x_2, \dots, \x_N)\rangle\\ = \langle\bigg((&D_N- 1)- \frac{\alpha Z}{|\x_N|}+ \underset{j= 1}{\overset{N- 1}{\sum}}\frac{\alpha}{|\x_j- \x_N|}\bigg)\\ &\times \phi(\x_1,\dots, \x_{N- 1})(\Lambda_+\psi_m)(\x_N), (\Lambda_+\psi_n)(\x_1)\phi(\x_2, \dots, \x_N)\rangle,
\end{split}\end{equation}
where $\mathcal{H}_{N-1}$ acts on the first $N-1$ electrons.

We introduce a cut--off function
\begin{equation}\label{chi again}
\chi\in C_0^\infty(\mathbb{R}^3), \quad \chi(\x)= \begin{cases}1, & |\x|\leqslant 1/4,\\ 0, & |\x|\geqslant 1/2,\end{cases} \quad \chi_n(\x):= \chi\Big(\frac{\x}{R_n}\Big).
\end{equation}

One has
\begin{equation}\begin{split}\label{kinetic2}
\bigg|\langle\bigg(&(D_N- 1)- \frac{\alpha Z}{|\x_N|}+ \underset{j= 1}{\overset{N- 1}{\sum}}\frac{\alpha}{|\x_j- \x_N|}\bigg)\\ &\times\phi(\x_1,\dots, \x_{N- 1})(\Lambda_+\psi_m)(\x_N), (\Lambda_+\psi_n)(\x_1)\phi(\x_2, \dots, \x_N)\rangle\bigg|\\ \leqslant \bigg|\langle\phi&(\x_1,\dots, \x_{N- 1})\bigg((D_N- 1)- \frac{\alpha Z}{|\x_N|}+ \underset{j= 1}{\overset{N- 1}{\sum}}\frac{\alpha}{|\x_j- \x_N|}\bigg)\\ &\times(\Lambda_+\psi_m)(\x_N), \chi_n(\x_1)(\Lambda_+\psi_n)(\x_1)\phi(\x_2, \dots, \x_N)\rangle\bigg|\\ + \bigg|\langle\big(1&- \chi_n(\x_1)\big)\phi(\x_1,\dots, \x_{N- 1})\bigg((D_N- 1)- \frac{\alpha Z}{|\x_N|}+ \underset{j= 1}{\overset{N- 1}{\sum}}\frac{\alpha}{|\x_j- \x_N|}\bigg)\\ &\times(\Lambda_+\psi_m)(\x_N), (\Lambda_+\psi_n)(\x_1)\phi(\x_2, \dots, \x_N)\rangle\bigg|\\ \leqslant \Big(\big\|&(D- 1)\Lambda_+\psi_m\big\|_{L_2(\mathbb{R}^3, \mathbb{C}^4)}+ 2\alpha (Z+ N- 1)\|\nabla\psi_m\|_{L_2(\mathbb{R}^3, \mathbb{C}^4)}\Big)\\ \times\bigg(\|\chi_n&\Lambda_+\psi_n\|_{L_2(\mathbb{R}^3, \mathbb{C}^4)}+ \Big\|\big(1- \chi_n(\x_1)\big)\phi(\x_1,\dots, \x_{N- 1})\Big\|_{L_2(\mathbb{R}^{3(N- 1)}, \mathbb{C}^{4^{(N- 1)}})}\bigg),
\end{split}\end{equation}
where for the electrostatic terms we have used the Hardy inequality, the fact that $\Lambda_+$ commutes with the gradient and the equality $\|\Lambda_+\|_{L_2(\mathbb{R}^3, \mathbb{C}^4)\rightarrow L_2(\mathbb{R}^3, \mathbb{C}^4)}= 1$.

Since only the first two components of $\psi_m$ are nonzero, we calculate
\begin{equation}\begin{split}\label{kinetic norm}
&\big\|(D- 1)\Lambda_+\psi_m\big\|^2\\ &= \langle(\al\cdot\mathbf{p}+ \beta- 1)(\al\cdot\mathbf{p}+ \beta- 1)\frac{\sqrt{|\mathbf{p}|^2+ 1}+ \al\cdot\mathbf{p}+ \beta}{2\sqrt{|\mathbf{p}|^2+ 1}}\hat\psi_m(\mathbf{p}), \hat\psi_m(\mathbf{p})\rangle\\ &= \langle|\mathbf{p}|^2\Big(\frac{1}{2}- \frac{1}{2\sqrt{|\mathbf{p}|^2+ 1}}\Big)\hat\psi_m(\mathbf{p}), \hat\psi_m(\mathbf{p})\rangle\\ &= \langle\frac{|\mathbf{q}|^2}{R_m^2}\Big(\frac{1}{2}- \frac{R_m}{2\sqrt{|\mathbf{q}|^2+ R_m^2}}\Big)\hat{\widetilde\psi}(\mathbf{q}), \hat{\widetilde\psi}(\mathbf{q})\rangle\leqslant \frac{1}{2R_m^2}\big\||\mathbf{q}|\hat{\widetilde\psi}(\mathbf{q})\big\|^2.
\end{split}\end{equation}
Thus the first factor on the \rhs of \eqref{kinetic2} decays as $\|\nabla\psi_m\|$, which is of order $R^{-1}$. 
The first term in the second factor of \eqref{kinetic2} is an exponentially decaying function of $R$ due to Lemma~\ref{decay lemma} and the support properties of $\psi_n$ and $\chi_n$. The second one also goes to zero as $R\rightarrow\infty$. We conclude that the second term on the \rhs of \eqref{expansion} decays faster than $R^{-1}$.

\paragraph{4.}

We proceed to the estimate of the last term on the \rhs of \eqref{expansion}. The key tool here is Lemma~\ref{decay lemma}.
We use the relation
\begin{equation}\begin{split}\label{n instead of m}
&\langle\widetilde{\mathcal{H}}_N(\phi\otimes\Lambda_+\psi_m), \phi\otimes\Lambda_+\psi_n\rangle\\& = \langle (D- 1)\Lambda_+\psi_m, \Lambda_+\psi_n\rangle- \langle\frac{\alpha Z}{|\x|}\Lambda_+\psi_m, \Lambda_+\psi_n\rangle\\ &+ \underset{i< N}{\sum}\langle\frac{\alpha}{|\x_i- \x_N|}\rho_\phi(\x_i)(\Lambda_+\psi_m)(\x_N), \rho_\phi(\x_i)(\Lambda_+\psi_n)(\x_N)\rangle.
\end{split}\end{equation}
For the kinetic energy term we have
\begin{equation}\begin{split}\label{again kinetic}
\big|\langle(D\Lambda_+- \Lambda_+)\psi_m, \Lambda_+\psi_n\rangle\big|&= \big|\langle\Lambda_+\psi_m, (D- 1)\psi_n\rangle\big|\\ &\leqslant \|\Lambda_+\psi_m\|_{L_2(\mathrm{supp}\,\psi_n)}\big\|(D- 1)\psi_n\big\|.
\end{split}\end{equation}
Notice that the norm $\big\|(D- 1)\psi_n\big\|$ is bounded uniformly in $n$ and $R$. 

Since
\[
\mathrm{supp}\,\psi_k\subset B\bigg(2^kR\Big(N- \frac{1}{5}\Big)\bigg)\setminus B\bigg(2^kR\Big(N- \frac{2}{5}\Big)\bigg), \quad k= 1, \dots, Q,
\]
Lemma~\ref{decay lemma} implies the exponential decay of $\|\Lambda_+\psi_m\|_{L_2(\mathrm{supp}\,\psi_n)}$, and hence of the \rhs of \eqref{again kinetic}, in $R$.

Let $B_{mn}$ be the ball with the radius $\frac{1}{2}(R_m+ R_m)$ centered at the origin.
For the interaction with the nucleus one has
\begin{equation}\begin{split}\label{in and out}
\langle &-\frac{\alpha Z}{|\x_N|}\Lambda_+\psi_m, \Lambda_+\psi_n\rangle\\ &= \alpha Z\bigg(\int\limits_{B_{mn}}+ \int\limits_{\mathbb{R}^3\setminus B_{mn}}\bigg)\frac{1}{|\x_N|}(\Lambda_+\psi_n)^*(\x_N)(\Lambda_+\psi_m)(\x_N)d\x_N.
\end{split}\end{equation}
Without loss of generality assume $n> m$. By Lemma~\ref{decay lemma}
\begin{equation}\label{in B_mn}
\|\Lambda_+\psi_n\|_{L_2(B_{mn})}\leqslant Ce^{-\delta R}
\end{equation}
for some $C> 0$ and $\delta> 0$ independent of $m$ and $n$. \\
By the Hardy inequality
\begin{equation}\label{by Hardy}
\Big\|\frac{1}{|\x_N|}\Lambda_+\psi_m\Big\|\leqslant 2\|\nabla\Lambda_+\psi_m\|\leqslant C_1|R_m|^{-1}.
\end{equation}
Together with \eqref{in B_mn} this implies
\begin{equation}\label{exp estimate}
\bigg|\int\limits_{B_{mn}}\frac{1}{|\x_N|}(\Lambda_+\psi_n)^*(\x_N)(\Lambda_+\psi_m)(\x_N)d\x_N\bigg|\leqslant Ce^{-\delta R}.
\end{equation}
Similarly, in $\mathbb{R}^3\setminus B_{mn}$ the estimates \eqref{in B_mn}, \eqref{by Hardy}, and \eqref{exp estimate} hold if we replace everywhere $m\leftrightarrow n$ and $B_{mn}$ by $\mathbb{R}^3\setminus B_{mn}$.

It follows that
\begin{equation}\label{nucleus term}
\Big|\langle\frac{\alpha Z}{|\x|}\Lambda_+\psi_m, \Lambda_+\psi_n\rangle\Big|\leqslant Ce^{-\delta R}.
\end{equation}

The last term in \eqref{n instead of m}, which describes the electron--electron interaction, can be estimated analogously to \eqref{in and out} --- \eqref{nucleus term}. Together with \eqref{again kinetic} and \eqref{nucleus term} this implies
\[
\big|\langle\widetilde{\mathcal{H}}_N(\phi\otimes\Lambda_+\psi_m), \phi\otimes\Lambda_+\psi_n\rangle\big|\leqslant Ce^{-\delta R},
\]
which completes the proof of Theorem~\ref{Z-theorem}.

\appendix

\section{Some Properties of the Bessel Functions $K_\nu$}\label{K-functions}

The modified Bessel (McDonald) functions are related to the Hankel functions by the formula
\[
K_\nu(z)= \frac{\pi}{2}e^{i\pi(\nu+ 1)/2}H_\nu^{(1)}(iz).
\]
These functions are positive and decaying for $z\in (0, \infty)$. Their asymptotics are (see \cite{GradshteynRyzhik2000} 8.446, 8.447.3, 8.451.6)
\begin{equation}\label{asymptotics}\begin{split}
K_\nu(z)&= \sqrt\frac{\pi}{2z}e^{-z}\bigg(1+ O\Big(\frac{1}{z}\Big)\bigg), \quad z\rightarrow+\infty;\\
K_0(z)&= -\log z\big(1+ o(1)\big), \quad K_1(z)= \frac{1}{z}\big(1+ o(1)\big), \quad z\rightarrow +0.
\end{split}\end{equation}
The derivatives of these functions are (see \cite{GradshteynRyzhik2000} 8.486.12, 8.486.18)
\begin{equation}\label{derivatives}
K_0'(z)= -K_1(z), \quad K_1'(z)= -K_0(z)- \frac{1}{z}K_1(z), \quad z\in (0, \infty).
\end{equation}

\section{Coordinate Representation for the Operator $\Lambda_+$}\label{integral formula}

\begin{lemma}\label{coordinate representation lemma}
Let $f\in L_2(\mathbb{R}^3, \mathbb{C}^4)\cap C_0^1(\mathbb{R}^3, \mathbb{C}^4)$ be a function in the coordinate representation. Then for $\Lambda_+f$ formula \eqref{Lambda_+ in configuration space} holds.
\end{lemma}
\begin{proof}{}
We start with the operator $2\Lambda_+- 1$, which is due to \eqref{Lambda_+} the multiplication by the matrix function $\frac{\D\al\cdot\mathbf{p}+ \beta}{\D\sqrt{|\mathbf{p}|^2+ 1}}$ in the momentum space. It can be factorized as $A\cdot B$ with
\[
A:= (\al\cdot\mathbf{p}+ \beta)\big(|\mathbf{p}|^2+ 1\big), \quad B:= \big(|\mathbf{p}|^2+ 1\big)^{-3/2}.
\]
In coordinate representation $B:L_2(\mathbb{R}^3, \mathbb{C}^4)\rightarrow H^3(\mathbb{R}^3, \mathbb{C}^4)$ is a bounded integral operator. Its kernel is given by the convergent integral
\[\begin{split}
B(\x, \y)&= \frac{1}{(2\pi)^3}\int\limits_{\mathbb{R}^3}\frac{e^{i\mathbf{p}\cdot(\x- \y)}}{\big(|\mathbf{p}|^2+ 1\big)^{3/2}}d\mathbf{p}\\ &= \frac{1}{2\pi^2}\int\limits_{0}^{\infty}\frac{p\,\sin\big(p|\x- \y|\big)}{|\x- \y|(p^2+ 1)^{3/2}}dp= \frac{1}{2\pi^2}K_0\big(|\x- \y|\big).
\end{split}\]
In configuration space $A$ is the differential operator $(-i\al\cdot\nabla+ \beta)(-\Delta+ 1)$ mapping $H^3(\mathbb{R}^3, \mathbb{C}^4)$ onto $L_2(\mathbb{R}^3, \mathbb{C}^4)$. Thus with the help of \eqref{derivatives} for any $f\in L_2(\mathbb{R}^3, \mathbb{C}^4)\cap C_0^1(\mathbb{R}^3, \mathbb{C}^4)$ we get
\begin{equation}\begin{split}\label{stage 1}
\big((2\Lambda_+- 1)f\big)(\x)= (-i\al\cdot\nabla+ \beta)(-\Delta+ 1)\frac{1}{2\pi^2}\int\limits_{\mathbb{R}^3}K_0\big(|\x- \y|\big)f(\y)d\y\\ = (-i\al\cdot\nabla+ \beta)\frac{1}{2\pi^2}\int\limits_{\mathbb{R}^3}\frac{K_1\big(|\x- \y|\big)}{|\x- \y|}f(\y)d\y\\ = (-i\al\cdot\nabla+ \beta)\frac{1}{2\pi^2}\int\limits_{\mathbb{R}^3}\frac{K_1\big(|\y|\big)}{|\y|}f(\x- \y)d\y.
\end{split}\end{equation}
The term with $\beta$ defines a function from $L_2(\mathbb{R}^3, \mathbb{C}^4)$, because $|\cdot|^{-1}K_1\big(|\cdot|\big)\in L_1(\mathbb{R}^3)$. We rewrite the gradient term on the \rhs of \eqref{stage 1} as
\begin{equation}\begin{split}\label{stage 2}
-i\al\cdot\nabla_\x\frac{1}{2\pi^2}&\int\limits_{\mathbb{R}^3}\frac{K_1\big(|\y|\big)}{|\y|}f(\x- \y)d\y\\&= \frac{i}{2\pi^2}\bigg(\int\limits_{\mathbb{R}^3\setminus B(\varepsilon)}+ \int\limits_{B(\varepsilon)}\bigg)\frac{K_1\big(|\y|\big)}{|\y|}\al\cdot\nabla_\y\big(f(\x- \y)\big)d\y.
\end{split}\end{equation}
The second integral on the \rhs of \eqref{stage 2} can be estimated as
\begin{equation}\label{in ball}
\bigg|\frac{i}{2\pi^2}\int\limits_{B(\varepsilon)}\frac{K_1\big(|\y|\big)}{|\y|}\al\cdot\nabla_\y\big(f(\x- \y)\big)d\y\bigg|\leqslant \frac{3}{2\pi^2}\|\nabla f\|_\infty\int\limits_{B(\varepsilon)}\frac{K_1\big(|\y|\big)}{|\y|}d\y,
\end{equation}
where the \rhs of \eqref{in ball} tends to zero as $\varepsilon\rightarrow 0$.
For the first integral on the \rhs of \eqref{stage 2} the integration by parts gives
\begin{equation}\begin{split}\label{stage 3}
&\frac{i}{2\pi^2}\int\limits_{\mathbb{R}^3\setminus B(\varepsilon)}\frac{K_1\big(|\y|\big)}{|\y|}\al\cdot\nabla_\y\big(f(\x- \y)\big)d\y\\ = &\frac{-i}{2\pi^2}\int\limits_{\mathbb{R}^3\setminus B(\varepsilon)}\!\!\al\cdot\nabla_\y\Big(\frac{K_1\big(|\y|\big)}{|\y|}\Big)f(\x- \y)d\y+ \frac{iK_1(\varepsilon)}{2\pi^2\varepsilon}\!\!\int\limits_{\partial B(\varepsilon)}\!\!\al\cdot\frac{\y}{|\y|}f(\x- \y)d\y.
\end{split}\end{equation}
We can rewrite the second integral on the \rhs of \eqref{stage 3} as the sum of two integrals, using the Taylor expansion $f(\x- \y)= f(\x)- \nabla f(\mathbf{z})\cdot\y$ with $\mathbf{z}$ lying on the segment connecting $\x$ and $\y$. The integral containing $f(\x)$ vanishes, because the function $\y|\y|^{-1}$ is odd.
The integral with $\nabla f(\mathbf{z})\y$ is different from zero only for $\x$ in the compact region 
\[
\Omega:= \big\{\x\big|\x\in\mathbb{R}^3, \: \mathrm{dist}\,\{\x, \, \mathrm{supp}\, f\}\leqslant \varepsilon\big\}.
\] 
For $\x\in\Omega$ we have
\begin{equation}\begin{split}\label{stage 5}
\bigg|\frac{iK_1(\varepsilon)}{2\pi^2\varepsilon}\int\limits_{\partial B(\varepsilon)}\Big(\al\cdot\frac{\y}{|\y|}\Big)\Big(\nabla f(\mathbf{z})\cdot\y\Big) d\y\bigg|\leqslant \frac{3K_1(\varepsilon)}{2\pi^2}\|\nabla f\|_\infty4\pi\varepsilon^2.
\end{split}\end{equation}
Hence the last term on the \rhs of \eqref{stage 3} converges to zero in the $L_2$--norm. Together with \eqref{stage 1} --- \eqref{stage 3} and \eqref{derivatives} this proves Lemma~\ref{coordinate representation lemma}.
\end{proof}

\section{Proof of Lemma \ref{lemma2}}\label{proof of the W lemma}

Let $f\in L_2(\mathbb{R}^{3N}), \: \mathrm{supp}\,f\subset [-2R, 2R]^{3N}$. Then
\[
f(\x)=\underset{\bk\in\mathbb{Z}_+^{3N}}{\sum}c_\bk\underset{i=1}{\overset{3N}{\prod}}\varphi_{k_i}(x_i),
\]
where
\[
\varphi_{k}(x)=\begin{cases}{\D\frac{1}{\sqrt{2R}}\sin\bigg(\pi k\Big(\frac{1}{2}+ \frac{x}{4R}\Big)\bigg),}& x\in [-2R, 2R],\\ 0,& x\notin [-2R, 2R],\end{cases}
\]
$c_\bk$ are the Fourier coefficients of $f\arrowvert_{[-2R, 2R]^{3N}}$. 

For the Fourier transform of $f$ we have
\begin{equation}\label{series}
\hat f(\mathbf{p})= \underset{\bk\in\mathbb{Z}_+^{3N}}{\sum}c_\bk\underset{i= 1}{\overset{3N}{\prod}}\hat\varphi_{k_i}(p_i),
\end{equation}
where
\begin{equation}\label{varphi}
 \hat\varphi_k(p)= \frac{4\sqrt{\pi R}ke^{i\frac{\pi}{2}(k- 1)}\sin\Big(\frac{\displaystyle\pi k}{\displaystyle 2}- 2pR\Big)}{\pi^2k^2- 16p^2R^2}.
\end{equation}
Let
\begin{equation}\label{second requirement}
L:= \frac{1536RNM}{\pi^3}+ 1.
\end{equation}
Assume that $f$ is orthogonal to the linear span of $L^{3N}$ functions 
\[
\bigg\{\underset{i= 1}{\overset{3N}{\prod}}\varphi_{k_i}(x_i), \quad k_i\in[0, L- 1]\cap\mathbb{Z}, \quad i= 1,\dots 3N\bigg\}.
\]
Then the summation in \eqref{series} can be restricted to
\[
\bk\in\underset{j= 1}{\overset{3N}{\bigcup}}\gamma_j, \quad \gamma_j:=\underset{l= 1}{\overset{j- 1}{\bigcap}}\{\bk\in\mathbb{Z}^{3N}_+\arrowvert k_l< L\}\cap\{\bk\in\mathbb{Z}_+^{3N}\arrowvert k_j\geqslant L\}.
\]
Obviously $\|\hat f\|_{L_2(\mathbb{R}^{3N})}^2= \underset{\bk\in\underset{j= 1}{\overset{3N}{\bigcup}}\gamma_j}{\sum}|c_\bk|^2$. 
On the other hand,
\[\begin{split}
&\|\hat f\|_{L_2(W_M)}\leqslant \underset{j= 1}{\overset{3N}{\sum}}\Big\|\underset{\bk\in\gamma_j}{\sum}c_\bk\underset{i= 1}{\overset{3N}{\prod}}\varphi_{k_i}(p_i)\Big\|_{L_2(\{|p_j|\leqslant M\})}\\&= \underset{j= 1}{\overset{3N}{\sum}}\bigg(\underset{\bk, \bk'\in\gamma_j}{\sum}\langle c_\bk, c_{\bk'}\rangle\int\limits_{-M}^{M}\varphi_{k_j}(p_j)\overline{\varphi_{k_j'}(p_j)}dp_j\underset{\substack{i= 1\\ i\neq j}}{\overset{3N}{\prod}}\int\limits_\mathbb{R}\varphi_{k_i}(p_i)\overline{\varphi_{k_i'}(p_i)}dp_i\bigg)^{1/2}\\&= \underset{j= 1}{\overset{3N}{\sum}}\bigg(\underset{\substack{k_i= 1\\ i< j}}{\overset{L}{\sum}}\underset{\substack{k_i= 1\\ i> j}}{\overset{\infty}{\sum}}\underset{k_j, k_j'= L}{\overset{\infty}{\sum}}\langle c_{(k_1,\dots,k_j,\dots,k_{3N})}, c_{(k_1,\dots,k_j',\dots,k_{3N})}\rangle\int\limits_{-M}^{M}\varphi_{k_j}(p)\overline{\varphi_{k_j'}(p)}dp\bigg)^{\frac{1}{2}}.
\end{split}\]
Since
\begin{equation}\label{first requirement}
k_j, \, k_j'\geqslant L> \frac{4\sqrt{2}MR}{\pi},
\end{equation}
we estimate
\[\begin{split}
\bigg|\int\limits_{-M}^{M}\varphi_{k_j}(p)\overline{\varphi_{k_j'}(p)}dp\bigg|\leqslant 16\pi Rk_jk_j'\int\limits_{-M}^M\frac{1}{|\pi^2k_j^2- 16p^2R^2||\pi^2k_j'^2- 16p^2R^2|}dp\\ \leqslant 16\pi Rk_jk_j'\cdot 2M\cdot\frac{2}{\pi^2k_j^2}\cdot\frac{2}{\pi^2k_j'^2}= \frac{128RM}{\pi^3k_jk_j'}.
\end{split}\]
Applying the Schwarz inequality, we arrive at
\[\begin{split}
\bigg|\underset{k_j, k_j'= L}{\overset{\infty}{\sum}}\langle c_{(k_1,\dots,k_j,\dots,k_{3N})}, c_{(k_1,\dots,k_j',\dots,k_{3N})}\rangle\int\limits_{-M}^{M}\varphi_{k_j}(p)\overline{\varphi_{k_j'}(p)}dp\bigg|\\ \leqslant \frac{128RM}{\pi^3}\underset{k_j'= L}{\overset{\infty}{\sum}}|c_{(k_1,\dots,k_j',\dots,k_{3N})}|^2\cdot\underset{k_j= L}{\overset{\infty}{\sum}}k_j^{-2}\\ \leqslant \frac{128RM}{\pi^3(L- 1)}\underset{k_j'= L}{\overset{\infty}{\sum}}|c_{(k_1,\dots,k_j',\dots,k_{3N})}|^2.
\end{split}\]
Therefore
\[\begin{split}
\|\hat f\|_{L_2(W_M)}&\leqslant \underset{j= 1}{\overset{3N}{\sum}}\bigg(\frac{128RM}{\pi^3(L- 1)}\underset{\bk\in\gamma_j}{\sum}|c_\bk|^2\bigg)^{1/2}\\ &\leqslant \sqrt{\frac{384RNM}{\pi^3(L- 1)}}\|\hat f\|_{L_2(\mathbb{R}^{3N})}= \frac{1}{2}\|\hat f\|_{L_2(\mathbb{R}^{3N})}.
\end{split}\]

\paragraph{Acknowledgement.} This work was supported by the EU Network "Analysis and Quantum" (HPRN--CT--2002--00277). S. M. was supported by the DFG grant SI 348/12--2, S. V. was supported by the DFG grant WE-1964/2.

\bibliographystyle{plain}
\bibliography{Bericht}

\noindent
Sergey Morozov\\
Mathematisches Institut der Ludwig--Maximilians--Universit\"at\\
Theresienstr. 39\\
D--80333 Munich, Germany\\
e--mail: morozov@mathematik.uni-muenchen.de\\

\noindent
Semjon Vugalter\\
Mathematisches Institut A, Universit\"at Stuttgart\\
Pfaffenwaldring 57\\
D--70569 Stuttgart, Germany\\
e--mail: wugalter@math.lmu.de

\end{document}